\documentclass[fleqn,usenatbib]{mnras}

\usepackage[T1]{fontenc}
\usepackage{ae,aecompl}

\usepackage{graphicx}    
\usepackage{amsmath}    
\usepackage{amssymb}    
\usepackage{stfloats}

\usepackage{soul,xcolor}

\title[Unsupervised Lens Finding]{Identifying Strong Lenses with Unsupervised Machine Learning using Convolutional Autoencoder}

\author[Ting-Yun Cheng et al.]{Ting-Yun Cheng,$^{1}$\thanks{E-mail:ting-yun.cheng@nottingham.ac.uk}
Nan Li,$^{2,1}$
Christopher J. Conselice,$^{1}$
Alfonso Arag\'on-Salamanca,$^{1}$
\newauthor
Simon Dye,$^{1}$
Robert B. Metcalf $^{3,4}$\\
\\
$^{1}$School of Physics and Astronomy, University of Nottingham, University Park, Nottingham, NG7 2RD, UK\\
$^{2}$CAS Key Laboratory of Space Astronomy and Technology, National Astronomical Observatories, Beijing 100012, People’s Republic of China\\
$^{3}$Dipartimento di Fisica \& Astronomia, Universit\'a di Bologna, Via Gobetti 93/2, 40129 Bologna, Italy\\
$^{4}$INAF-Osservatorio Astronomico di Bologna, Via Ranzani 1, 40127 Bologna, Italy\\
}

\date{Accepted 2020 April 8. Received 2020 April 8; in original form 2019 November 6.}

\pubyear{2019}

\begin{document}
\setstcolor{red}
\setulcolor{blue}
\label{firstpage}
\pagerange{\pageref{firstpage}--\pageref{lastpage}}
\maketitle

\begin{abstract}
In this paper we develop a new unsupervised machine learning technique comprised of a feature extractor, a convolutional autoencoder (CAE), and a clustering algorithm consisting of a Bayesian Gaussian mixture model (BGM).  We apply this technique to visual band space-based simulated imaging data from the Euclid Space Telescope using data from the Strong Gravitational Lenses Finding Challenge. Our technique promisingly captures a variety of lensing features such as Einstein rings with different radii, distorted arc structures, etc, without using predefined labels.  After the clustering process, we obtain several classification clusters separated by different visual features which are seen in the images. Our method successfully picks up $\sim$63\ percent of lensing images from all lenses in the training set. With the assumed probability proposed in this study, this technique reaches an accuracy of $77.25\pm 0.48$\% in binary classification using the training set. Additionally, our unsupervised clustering process can be used as the preliminary classification for future surveys of lenses to efficiently select targets and to speed up the labelling process. As the starting point of the astronomical application using this technique, we not only explore the application to gravitationally lensed systems, but also discuss the limitations and potential future uses of this technique.
\end{abstract}


\section{Introduction}

Gravitational lensing has become established as a powerful probe in many areas of astrophysics and cosmology \citep[e.g., see reviews by][and references therein]{Mao2012,Meneghetti2013,Fu2014,Rahvar2015,Mandelbaum2018,Bartelmann2017}. The phenomenon has been detected since \citet{Walsh1979} and over a wide range of scales, from Mpc in the weak-lensing regime \citep[e.g.][]{Bacon2000, Hamana2003,Castro2005,Schmidt2008,Bernardeau2012,Jee2016,Kilbinger2017,Troxel2018}, to kpc in strong lensing \citep[e.g.][]{Lynds1986,Soucail1987,Fort1988,Hudson1998,Hewitt1988,Barvainis2002,Oldham2017,Stacey2018,Talbot2018} and down to pc and sub-pc scales probed by microlensing \citep[e.g.][]{Bruce2017,Shvartzvald2017,Han2018}. As such, lensing can be exploited to measure the distribution of mass in the Universe \citep[e.g.][]{Newman2013,Han2015,Diego2018,Jauzac2018}, enhance the study of lensed high redshift galaxies \citep[e.g.][]{Coe2013,Jones2013,Stark2015, Dye2015} and constrain cosmological models \citep[e.g.][]{Suyu2013,Suyu2014,Liao2015,Magana2015}, amongst other applications.

Galaxy-galaxy strong lensing (GGSL) is a particular case of gravitational lensing in which the background source and foreground lens are both galaxies, and the lensing effect is sufficient to distort images of the source into arcs or even Einstein rings. Since the discovery of the first GGSL system in 1988 \citep{Hewitt1988}, many valuable scientific applications have been realized for them, such as studying galaxy mass density profiles \citep[e.g.][]{Sonnenfeld2015,Shu2016b,Kung2018}, detecting galaxy substructure \citep[e.g.][]{Vegetti2014,Hezaveh2016,Bayer2018}, measuring cosmological parameters \citep[e.g][]{Collett2014,Rana2017,Suyu2017}, investigating the nature of high redshift sources \citep{Bayliss2017,Dye2018,Sharda2018}, and constraining the properties of the self-interaction physics of dark matter \citep[e.g.][]{Shu2016,Gilman2018,Kummer2018}.

Increasing the statistical power of these applications and improving sample uniformity requires a large increase in the number of known GGSL systems. Next generation imaging surveys arising from facilities such as Euclid, the Large Synoptic Survey Telescope (LSST), and the Wide Field Infrared Survey Telescope (WFIRST) are anticipated to increase the number of known GGSLs by several orders of magnitude \citep{Collett2015}. These forthcoming datasets present a challenge for identifying new GGSLs using automated procedures that operate in an efficient and reliable manner. To this end, a number of algorithms have been developed to detect GGSLs in image data by recognising arc-like features and Einstein rings \citep[e.g.][]{Gavazzi2014,Joseph2014,Paraficz2016,Bom2017}. In addition, instead of recognising arc-like features, an alternative detection technique that has had some success is to attempt to fit lens mass models to candidate GGSLs and reject those systems that do not converge \citep{Marshall2009,Sonnenfeld2018}. 

More recently, efforts to automate GGSL finding have turned to machine learning algorithms given their strong performance in the general field of image recognition. In particular, a class of deep learning networks known as convolutional neural networks (CNNs) can be trained to identify specific image features and thereby distinguish different categories of objects. In astronomy, these algorithms are beginning to be used in categorizing galaxy morphologies \citep[e.g.][]{Dieleman2015, Huertas-Company2015, Domnguez2018, Cheng2019}, measuring photometric redshifts \citep{Cavuoti2017, Sadeh2016, Samui2017}, and classifying supernovae \citep{Lochner2016}. Recent work has also shown that CNNs can be used to perform lens modelling as a vastly more efficient alternative to traditional parametric methods \citep{Hezaveh2017, James2019}.

The application of CNNs for detecting these GGSL systems has reached a high success rate in binary classification \citep{Jacobs2017, Petrillo2017, Ostrovski2017, Bom2017, Hartley2017, Avestruz2017, Lanusse2018}; however, the application of supervised machine learning such as CNNs is prone to human bias and training set bias which may not properly represent the diversity of real GGSL systems observed in future surveys. Additionally, GGSLs are rare events in the Universe so that there is insufficiently homogeneous data for training in supervised machine learning methods. Although simulated images can be used for training, they are generally lacking in the complexity of real observed data.

Unlike supervised machine learning which requires a large amount of labelled data, which can be expensive and misleading, unsupervised machine learning can be applied directly to observed data without labelling that helps to reduce human bias while training a machine. Therefore, scientists have started to explore the application of unsupervised machine learning to, e.g. phtometric redshifts \citep{Geach2012, Way2012, Carrasco2014, Siudek2018}, as well as to classification using photometry or spectroscopy \citep{D'Abrusco2012, Fustes2013, Siudek2018b}.

The application of unsupervised machine learning becomes more challenging when using high dimensional data such as images. \citet{Hocking2018} and \citet{Martin2019} are amongst the first studies of unsupervised machine learning applications using imaging data and who applied the Growing Neural Gas algorithm \citep{Fritzke1995}. In our study, we explore a different technique from \citet{Hocking2018} and \citet{Martin2019} in which we apply a convolutional autoencoder (CAE) \citep{Masci2011} to do feature extraction before connecting with unsupervised machine learning algorithms.

Our unsupervised machine learning gives an alternative way to approach human identifications without labels on automate GGSL detection that can be also used as the preliminary selection in future surveys to find initial set of lenses. Furthermore, without human bias, we can explore unique GGSL systems that would not be found by other methods without this unsupervised machine learning technique.

This paper is structured as follows. The unsupervised machine learning technique adopted in this paper is introduced in Section~{\ref{sec_method}}. Details about the implementation, including the pipeline and dataset, are described in Section~{\ref{sec_implement}}. Section~{\ref{sec_results}} discusses our findings. The discussion of future work is discussed in Section~\ref{sec_discussion}. Finally, the conclusions are presented in Section~{\ref{sec_conclusion}}.

\section{Methodology}
\label{sec_method}
The application of unsupervised machine learning has achieved successes on one dimensional data in astronomy such as with spectroscopic data or photometric parameters \citep[e.g.][]{D'Abrusco2012, Geach2012, Way2012, Fustes2013, Carrasco2014, Siudek2018, Siudek2018b}. However, the capability of unsupervised machine learning for high dimensional data such as imaging data has not been well explored.

The latest astronomical approaches of unsupervised machine learning application using imaging data made by \citet{Hocking2018} and \citet{Martin2019} apply the concept of deep clustering. Deep clustering \citep[e.g.][]{Hsu2015, Hershey2015, Xie2016, Caron2018} is a clustering method that groups together the features learned through a neural network. Both \citet{Hocking2018} and \citet{Martin2019} apply a neural network called `growing neural gas algorithm (GNG)' \citep{Fritzke1995}, which is a type of self-organizing map (Kohonen map) \citep{Kohonen1997}, to create feature maps from imaging data. They then connect these feature maps with a hierarchical clustering technique \citep{Hastie2009}.

In addition to neural networks, studies in computer science also use an architecture of both supervised (CNNs) and unsupervised convolutional neural networks (UCNNs) \citep[e.g.][]{Dosovitskiy2014} to the process of feature learning \citep[computer science: e.g.][]{Dundar2015, Bautista2016, Borji2017}.

There are a variety of unsupervised approaching for deep clustering using the architecture of CNNs. However, most of them use alternative unsupervised algorithms (e.g. k-mean) to calculate the weights between layers that reduces the power of CNNs for capturing features fit with human judgement when using imaging data. Therefore, instead of variational CNNs, we propose to use a convolutional autoencoder (CAE, Section~\ref{sec_cae}) as the feature extractor \citep{Masci2011} in this study. This preserves the intrinsic features of the images \citep{Guo2017, Li2017, Dizaji2017}. For the clustering part we apply the Bayesian Gaussian mixture model (BGM, Section~\ref{sec_uml}) to images presented by the features extracted by the CAE to group the input features in a high-dimensional feature space.
\subsection{Convolutional AutoEncoder (CAE)}
\label{sec_cae}
\begin{figure*}
\begin{center}
\graphicspath{}
	\includegraphics[width=2.1\columnwidth]{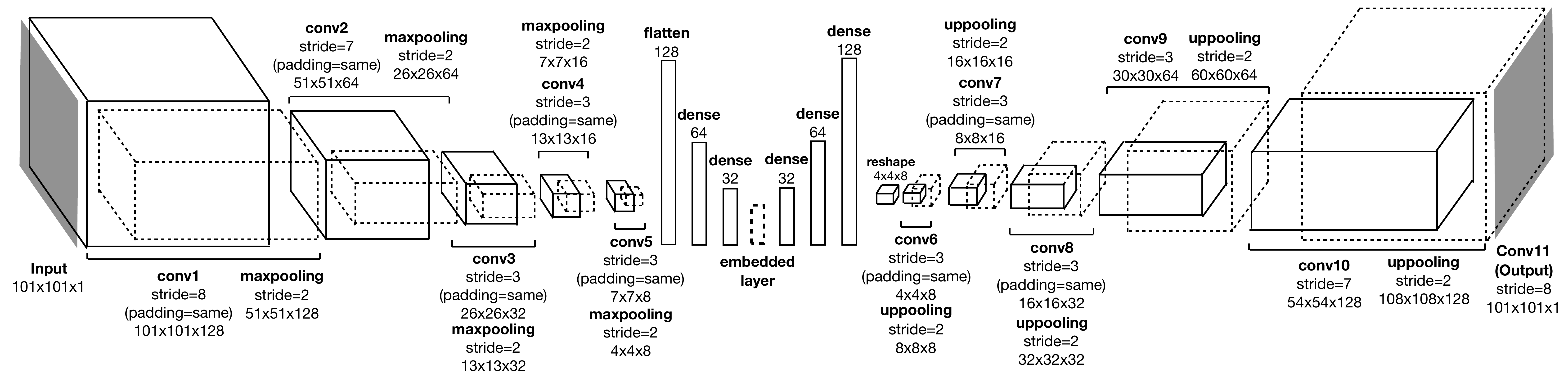}
   	\caption{The schematic overview for the architecture of our convolutional autoencoder (CAE) which is composed of two parts, the encoder and the decoder. The encoder starts from an input image with a size of 101 by 101 pixels (leftmost side) which is then connected with 5 convolutional layers (filter size: 128, 64, 32, 16, and 8). Each convolutional layer is followed a pooling layer. Three dense layers (units: 128, 64, 32) follow the fifth convolutional layer. The central dense layer of the architecture is called the `embedded layer'. We explore different number of units for this layer in this study (section~\ref{sec_training}). The decoder has similar structure to the encoder, and we use the units in the embedded layer to reproduce the input image as the output (rightmost side).}
    \label{fig:cae}
\end{center}
\end{figure*}

The convolutional autoencoder (CAE) \citep{Masci2011} is a kind of autoencoder (AE) which is mostly well known for denoising images \citep{Vincent2010}. The function of an AE is to learn a prior which features best represent the data distribution. With a limited number of features available, an AE intentionally captures significant features from images rather than the details of the background noise. The AE can then reconstruct images with this obtained prior.

The CAE improves the performance of an AE by considering the structures within two dimensional images that are ignored in the AE. Hence, the CAE preserves spatially localised features from image patches, while the AE can only obtain the global features.

The architecture of the CAE used in this study is shown in Fig.~\ref{fig:cae}. It includes two parts: encoder (left) and decoder (right). The encoder extracts the representative features from the input image. For an input $x$, the $j$-th representative feature map is given by
\begin{equation}
    	{ h }^{ j }=f\left( x\ast { W }^{ j }+{ b }^{ j } \right), 
\end{equation}
where $W$ are filters, $\ast$ denotes the 2 dimensional convolution operation, $b$ is the corresponding bias of the $j$-th feature map, and $f$ is an activation function. The encoder in this study is built with five convolutional layers (filter size: 128, 64, 32, 16, and 8) and three dense layers (units: 128, 64. 32). The activation function used in the convolutional layers is the Rectified Linear Unit ({\texttt{ReLu}}) \citep{Nair2010} such that $f(z)=0$ if $z<0$ while $f(z)=z$ if $z\ge0$. Each convolutional layer is followed by a pooling layer with a size of 2 by 2 pixels. The pooling layer is also referred to as a downsampling layer which is to reduce the spatial size and reduce the parameters involved in the CAE. 

The decoder then reproduces input images from the representative features; therefore, the architecture of the decoder is symmetric but reverse to that of the encoder. We invert the procedure of the encoder to reconstruct the representative feature maps back to the original shape of the input image by using the following formula:
\begin{equation}
\label{equ:reconstruction}
    	{ y }=f\left( \sum _{ j\in H }^{  }{ { h }^{ j }\ast { \overset { \sim  }{ W }  }^{ j } } +c \right) , 
\end{equation}
where ${ \overset { \sim  }{ W }}$ is the flip operator that transposes the weights, $\ast$ denotes 2 dimensional convolution operation, $c$ is the corresponding bias, $f$ is an activation function, and $H$ indicates the group of feature maps. The design for the number of filters in the convolution processes is based on the size of input images to form a symmetric structure between encoder and decoder. 

We have three dense layers (units: 32, 64, and 128), five convolutional layers (filter sizes: 8, 16, 32, 64, and 128) using the {\texttt{ReLu}} activation function \citep{Nair2010}, and an extra convolutional layer (filter: 1) using the {\texttt{softmax}} function \citep{Bishop2006}, $f\left( { z } \right) ={ exp\left( z \right)  }/{ \sum { exp\left( { z }^{ j } \right)  }  }$, as the output for the decoder. Each convolutional layer apart from the last layer (output) is followed with an upsampling layer which has the opposite function to the pooling layer that is used for recovering the resolution. 

The central dense layer of the CAE is called the `embedded layer (EL)' (see Fig.~\ref{fig:cae}). This is composed of the final latent representation features used for the reconstruction of the input images. In section~\ref{sec_training}, we explore the number of units required for the EL. 

The CAE extracts the latent representative feature maps by minimizing the reconstruction error. In this study, we use {\texttt{binary{\_}crossentropy}} in the {\sc{keras}} library\footnote{https://keras.io} to calculate the loss function of the CAE which is given by the following form, 
\begin{equation}
    	L=-\frac { 1 }{ N } \sum _{ n=1 }^{ N }{ \left[ { y }^{ n }\log { { \hat { y }  }^{ n }+\left( 1-{ y }^{ n } \right) \log { \left( 1-{ \hat { y }  }^{ n } \right)  }  }  \right]  },
\end{equation}
where $N$ is the number of samples, ${ y }^{ n }$ are targets, and ${ \hat { y }  }^{ n }$ are the reconstructed images (equation~\ref{equ:reconstruction}). We build our CAE using the {\sc{keras}} library and the {\sc{TensorFlow}} backend \footnote{https://www.tensorflow.org} \citep{tensorflow2015}.

\subsection{Bayesian Gaussian Mixture Model (BGM)}
\label{sec_uml}
A Gaussian mixture model is a probabilistic model for either density estimation or clustering using a mixture of a finite number of Gaussian distributions to describe the distributions of data points on a feature map. Given $K$ components, the algorithm uses {\texttt{Kmeans}} to initialise the weights, the means, and the covariances for the $K$ Gaussian distributions which are given in the form:

\begin{equation}
\label{equ:bgm}
    	p\left( x \right) =\sum _{ k=1 }^{ K }{ { w }_{ k }G } \left( { x }|{ { u }_{ k },{ \varepsilon  }_{ k } } \right),
\end{equation}
where $G \left( { x }|{ { u }_{ k },{ \varepsilon  }_{ k } } \right) $ represents $k$-th Gaussian, ${ u }_{ k }$ denotes the mean of the $k$-th Gaussian distribution, ${ \varepsilon  }_{ k }$ is the covariance matrix of the $k$-th Gaussian, and ${ w }_{ k }$ is the prior probability (weight) of the $k$-th Gaussian where,
\begin{equation}
\label{equ:sum_proba_1}
    	\sum _{ k=1 }^{ K }{ { w }_{ k } } =1.
\end{equation}
\noindent The algorithm then searches for the best fit of the $K$ Gaussian distributions to the data distribution through an iterative process.

\begin{figure}
\begin{center}
\graphicspath{}
	\includegraphics[width=\columnwidth]{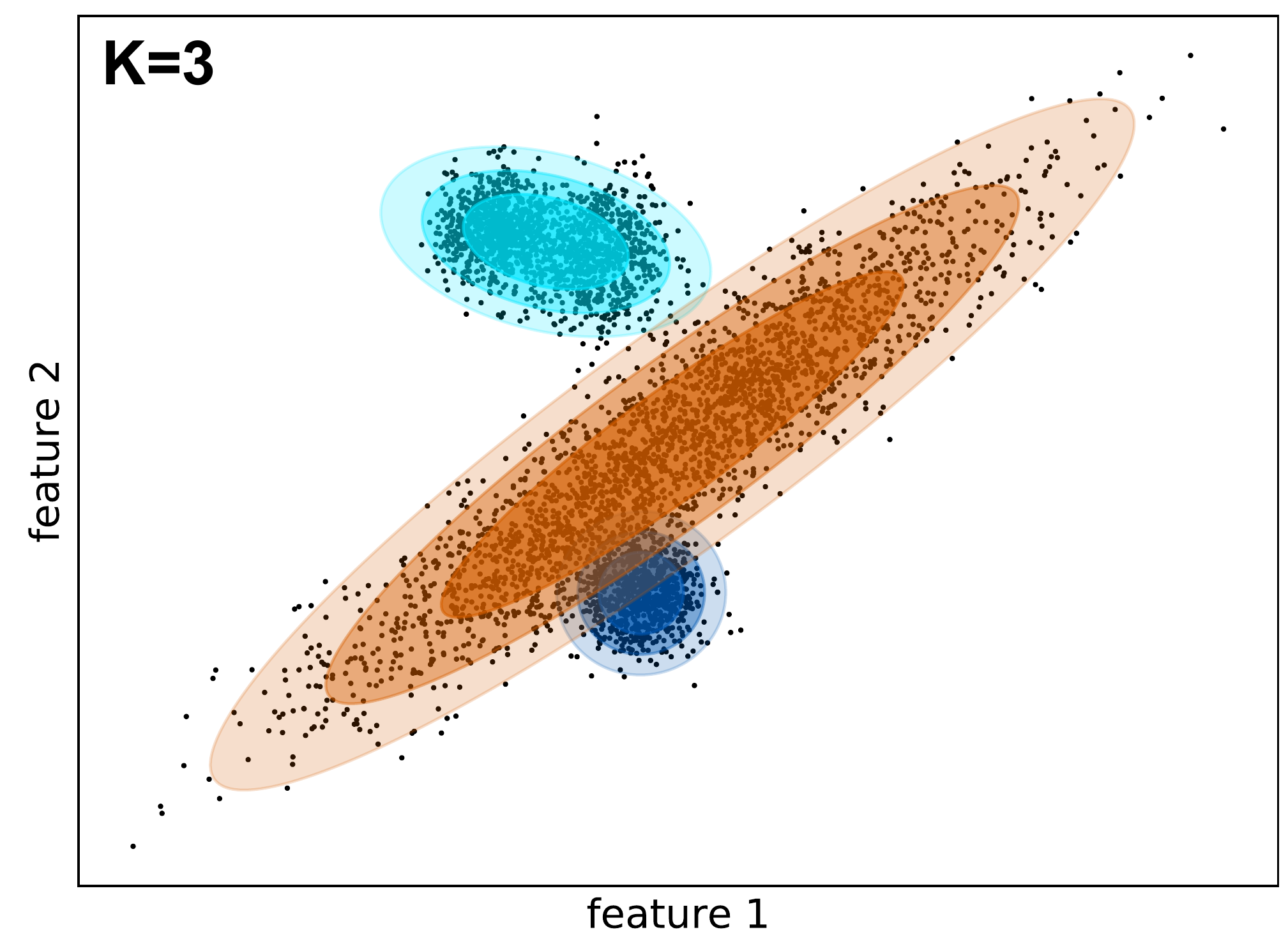}
   	\caption{An illustration of the Gaussian Mixture model we use. The $K$ value is the number of Gaussian distributions. The black dots show the data distribution on the feature map, and the coloured ellipses represent the three Gaussian distribution we applied here to fit the data distribution.}
    \label{fig:show_bgm}
\end{center}
\end{figure}

A two dimensional illustration of the BGM is shown in Fig.~\ref{fig:show_bgm} (Equation~\ref{equ:bgm}). The input data are distributed on the feature map (black dots). We use 3 Gaussian distributions in this illustration (coloured ellipses), to fit the data distribution on the feature map. 

In unsupervised learning, expectation-maximization (EM) \citep{Hartley1958, Dempster1977, McLachlan1997} is used to find the maximal log-likelihood estimates for the parameters of the Gaussian mixture model by an iterative process. The log-likelihood of the Gaussian mixture model is calculated using the formula:
\begin{equation}
    	\ln { \left[ p\left( { x }|{ u,\varepsilon ,w } \right)  \right]  } =\sum _{ n=1 }^{ N }{ \left\{ \ln { \left[ \sum _{ k=1 }^{ K }{ { w }_{ k }G } \left( { x }|{ { u }_{ k },{ \varepsilon  }_{ k } } \right)  \right]  }  \right\}  },
\end{equation}
where $N$ is the number of samples.

The Bayesian Gaussian mixture model (BGM) is a variational Gaussian mixture model \citep{Kullback1951, Attias2000, Bishop2006} which maximises the evidence lower bound (ELBO) \citep{Kullback1951} in the log-likelihood. In this study, we apply the BGM from the {\sc{scikit-learn}} library \footnote{https://scikit-learn.org/stable/index.html} \citep{Pedregosa2011}.

\section{Implementation}
\label{sec_implement}
In this section, we first introduce the datasets used in this study. The feature learning procedure is discussed in section~\ref{sec_training}. Section~{\ref{sec_clf}} presents the clustering and classifying phase which explains how to obtain the predicted lensing probability for each image. The tests for quantifying the performance of the classifications are described in section~{\ref{sec_exam}}.
\subsection{Data Sets}
\label{sec_dataset}
The strong lensing data are from the Strong Gravitational Lens Finding Challenge (Lens Finding Challenge) \citep{Metcalf2018}. The generation of mock images follows the procedures described in \citet{Grazian2004} and \citet{Meneghetti2008}, and starts with a cosmological N-boby simulation, the Millennium simulation \citep{Boylan-Kolchin2009}. The background objects are modeled by the sources from the Hubble Ultra Deep Field (UDF). The detail of the simulation setup can be found in \citet{Metcalf2018}.

We use the datasets which mimic the data quality of observations that will be taken by the Euclid Space Telescope \citep{Laureijs2011} in the visual (VIS) band. The pixel size is set to 0.1 arcsec and a Gaussian point spread function is applied to the images. Additionally, the noise follows a Gaussian distribution which is added to the final images \citep{Metcalf2018}.

\begin{figure*}
\begin{center}
\graphicspath{}
	\includegraphics[width=2\columnwidth]{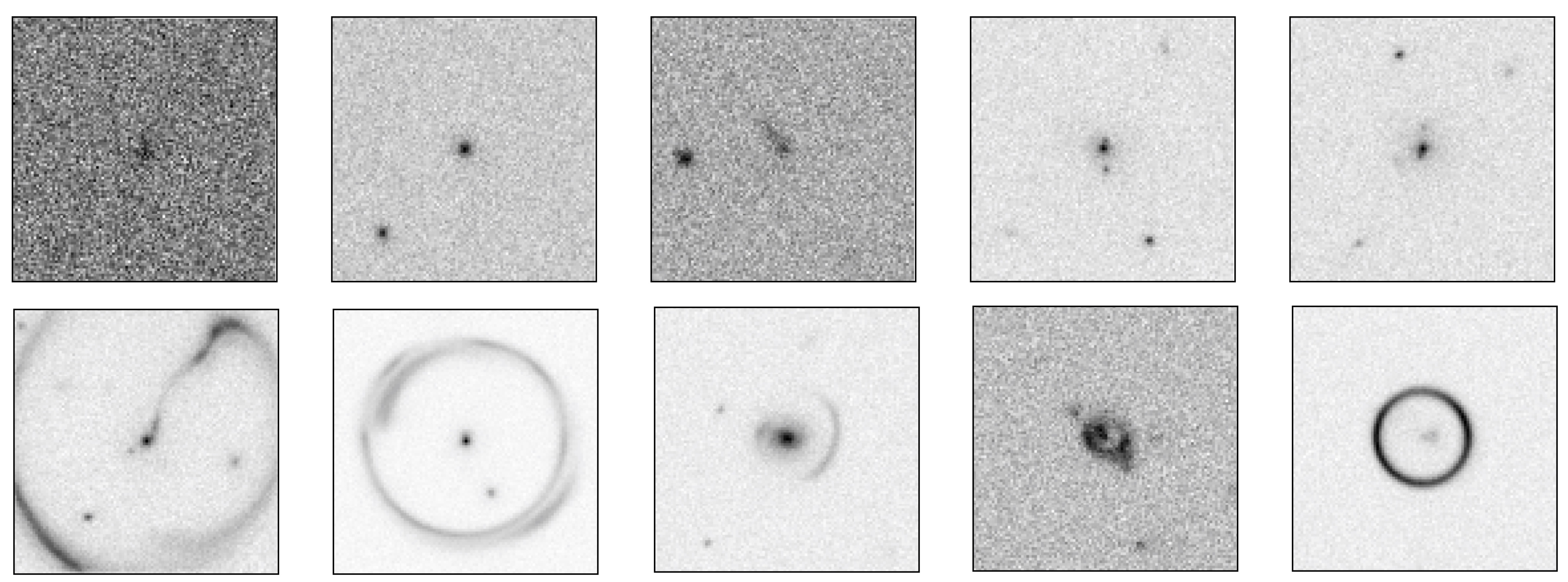}
   	\caption{An example of the training set for Lens Finding Challenge {\it Top:} non-lensing image; {\it Bottom:} lensing image.}
    \label{fig:examples_trainingdata}
\end{center}
\end{figure*}

There are 20,000 labelled images with lenses for training (13,968 lensing images; 6,032 non-lensing images, see Fig~\ref{fig:examples_trainingdata}) and 100,000 unlabelled images with lenses for testing in the Lens Finding Challenge.

We split the training set received from the Lens Finding Challenge into two parts, our own training set and testing sets. We randomly pick 12,800 lensing images out of 13,968 lensing images to obtain enough information for feature extraction. Additionally, we rotate a random set of 3,200 non-lensing images 4 times (0, 90, 180, 270 degrees)
to obtain the same number of images as there are lensing images (12,800 images) for our training set. An extra insignificant Gaussian noise is added into the rotated images to enhance the difference between the rotated images and the original images. The ratio between lensing and non-lensing images is 1 in the training set to make the convolutional autoencoder (CAE) consider both types equally when extracting features.

The rest of the images are the candidates for the testing sets. In our own testing sets, we initially have 1,168 lensing and 2,832 non-lensing images, which are leftover from the selection of the training set. We rotate the non-lensing images 4 times (0, 90, 180, 270 degrees) and add Gaussian noise to increase the number of images to 11,328 non-lensing images.

We test several different ratios between the number of lensing and non-lensing images to mimic a more realistic case. To avoid a biased influence from lensing images, we use the same set of lensing images in the testing process. We generate different ratios by randomly and repeatedly picking samples from the set of rotated non-lensing images. The arrangement is shown in Table~\ref{tab:testing} and is based on the prediction of \citet{Collett2015} which forecasts 2,400, 120,000, and 170,000 detectable galaxy-galaxy strong lenses out of 11 million lenses from their model for lensing systems in the Dark Energy Survey\footnote{https://www.darkenergysurvey.org/}, Large Synoptic Survey Telescope\footnote{https://www.lsst.org}, and Euclid Space Telescope, respectively. This arrangement for the fractions of lensing images in the testing sets cover from 50 percent to 0.01 percent.
\begin{table}
	\centering
	\begin{tabular}{ccc} 
		\hline
		Labels & Ratios & Number of data in each type\\
		\hline\hline
		\multicolumn{1}{c|}{1} &\multicolumn{1}{l}{1:1} &\multicolumn{1}{l}{lensing:1168/ non-lensing:1168}\\
        \multicolumn{1}{c|}{2} &\multicolumn{1}{l}{1:2} &\multicolumn{1}{l}{lensing:1168/ non-lensing:2336}\\
        \multicolumn{1}{c|}{3} &\multicolumn{1}{l}{1:20} &\multicolumn{1}{l}{lensing:1168/ non-lensing:23360}\\
        \multicolumn{1}{c|}{4} &\multicolumn{1}{l}{1:50} &\multicolumn{1}{l}{lensing:1168/ non-lensing:58400}\\
        \multicolumn{1}{c|}{5} &\multicolumn{1}{l}{1:100} &\multicolumn{1}{l}{lensing:1168/ non-lensing:116800}\\
        \multicolumn{1}{c|}{6} &\multicolumn{1}{l}{1:1000} &\multicolumn{1}{l}{lensing:1168/ non-lensing:1168000}\\
        \multicolumn{1}{c|}{7} &\multicolumn{1}{l}{1:10000} &\multicolumn{1}{l}{lensing:1168/ non-lensing:11680000}\\
		\hline
	\end{tabular}
	\caption{The arrangement of the testing datasets in this study. The ratios between lensing and non-lensing images are shown in the second column and the content included in the datasets are shown in the third column.}
	\label{tab:testing}
\end{table}
\subsection{Feature Learning}
\label{sec_training}
\begin{figure}
\begin{center}
\graphicspath{}
	\includegraphics[width=\columnwidth]{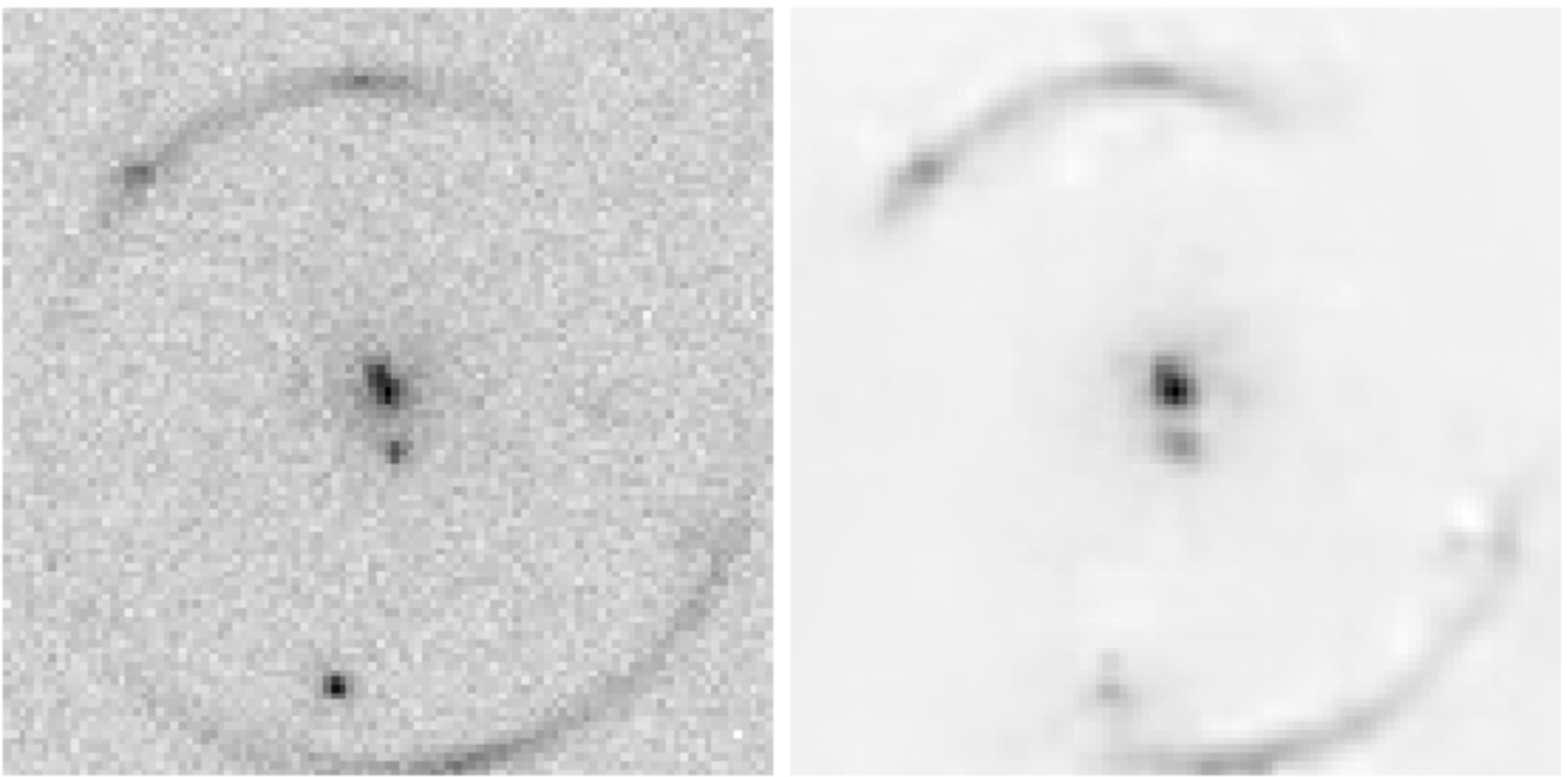}
   	\caption{An example of the denosing process. {\it Left}: the original image. {\it Right}: the image after denoising by an alternative CAE architecture described in section~\ref{sec_training}}
    \label{fig:denoise}
\end{center}
\end{figure}

There are three steps to take in the application of the techniques used in this study: (1) denoising the images by the convolutional autoencoder (CAE) with a simpler structure; (2) extracting the features of the images using the CAE (Fig.~\ref{fig:cae}); (3) identifying clusters using the features extracted from the CAE by the Bayesian Gaussian mixture model (BGM).

We recognise that the background noise in images influences the result of feature extraction because the CAE can overfit to the noise. As mentioned in Section~\ref{sec_cae}, an autoencoder learns the prior distribution from the input images (with noise) which preferentially captures the representatively strong features in images, but ignores insignificant features such as noise. Therefore, the reconstruction based on the prior distribution learnt through an autoencoder generates noiseless reconstructed images. We apply a CAE with a simpler architecture without hidden layers in Fig.~\ref{fig:cae} to generate noiseless images at the first step. 

This architecture contains five convolutional layers (filters: 128, 64, 32, 16, 8) with {\texttt{ReLu}} activation function for the encoder, five convolutional layers (filters: 8, 16, 32, 64, 128) with {\texttt{ReLu}} activation function for the decoder, an output layer with a {\texttt{softmax}} activation function. Each convolutional layer is followed with either a pooling layer or an upsampling layer in the encoder or decoder, respectively. The effect is shown in Fig.~\ref{fig:denoise}. The left panel is the original image, and the right panel is the image after denoising. Although the reconstructed images have lower resolution, they preserve and emphasize the features of lenses and sources that helps our CAE (Fig.~\ref{fig:cae}) to capture meaningfully representative features from images in the second step.

Secondly, we apply the CAE to carry out feature extraction (Fig.~\ref{fig:cae}). The final representative features are located within the embedded layer (EL) in the centre of the architecture. Finally, these extracted features are the input for the third step - clustering using the Bayesian Gaussian mixture model (BGM) utilising the representative features extracted by the CAE from the images. 

The number of clusters, $K$, when using unsupervised machine learning is generally unknown and difficult to be determined as there is not yet a reliable optimisation process to decide this quantity in unsupervised machine learning.

In \citet{Guo2017}, they suggest the number of extracted features to use should be the same as the number of clusters of datasets used (MNIST\footnote{http://yann.lecun.com/exdb/mnist/}). These number of clusters are however known in their case. This arrangement ensures that: (1) the dimension of the embedded layer was lower than the input data, and (2) the network could be trained directly in an end-to-end manner without any regularisations.

In contrast, the number of clusters is unknown in our work, and the number of extracted features is a hyper-parameter which can be controlled. Therefore, we decided to set the number of clusters, $K$, using the opposite concept from \citet{Guo2017}, to be the same as the number of extracted features. 

We can explain this decision using a simplified condition by assuming each feature decides one cluster; therefore, the number of features would be the intrinsic minimal number of clusters used.

The process of feature learning using the CAE is computationally expensive. Presently, it takes up to 5 days to train 100,000 images running on a NVIDIA GeForce GTX 1080 Ti GPU. In the future a more complex analysis of this issue can be carried out once computing power significantly improves.
\subsection{Clustering and classifying}
\label{sec_clf}
After clustering by the Bayesian Gaussian mixture model (BGM), we obtain the probability of each image belonging to each cluster. These probabilities are used to calculate the overall probability of each image being a strong lensing system.

With the probability of the $n$-th image to the $k$-th cluster, given by ${ P }^{ kn } $ and known fractions of lensing and non-lensing images in the $k$-th cluster, ${ P }_{ len }^{ k }$ and ${ P }_{ non }^{ k }$, we are able to calculate the predicted probability of different types, lensing (${P}_{len}^{ n }$) and non-lensing (${P}_{non}^{ n }$) for the $n$-th image by the formulas:
\begin{equation}
        \label{equ:probability}
    	\begin{cases} { P }_{ len }^{ n }=\sum _{ k=1 }^{ K }{ { P }_{ len }^{ k }{ \times P }^{ kn } }  \\ { P }_{ non }^{ n }=\sum _{ k=1 }^{ K }{ { P }_{ non }^{ k }{ \times P }^{ kn } }  \end{cases}.
\end{equation}

However, our technique is meant to be unsupervised; therefore, ${ P }_{ len }^{ k }$ and ${ P }_{ non }^{ k }$ are unknown. Without the label information, the network has no prior knowledge regarding classes of lensing or non-lensing. Therefore, to be able to compare the performance of this work and others, we must involve human classification after the step of the feature learning.

\begin{figure}
\begin{center}
\graphicspath{}
	\includegraphics[width=0.95\columnwidth]{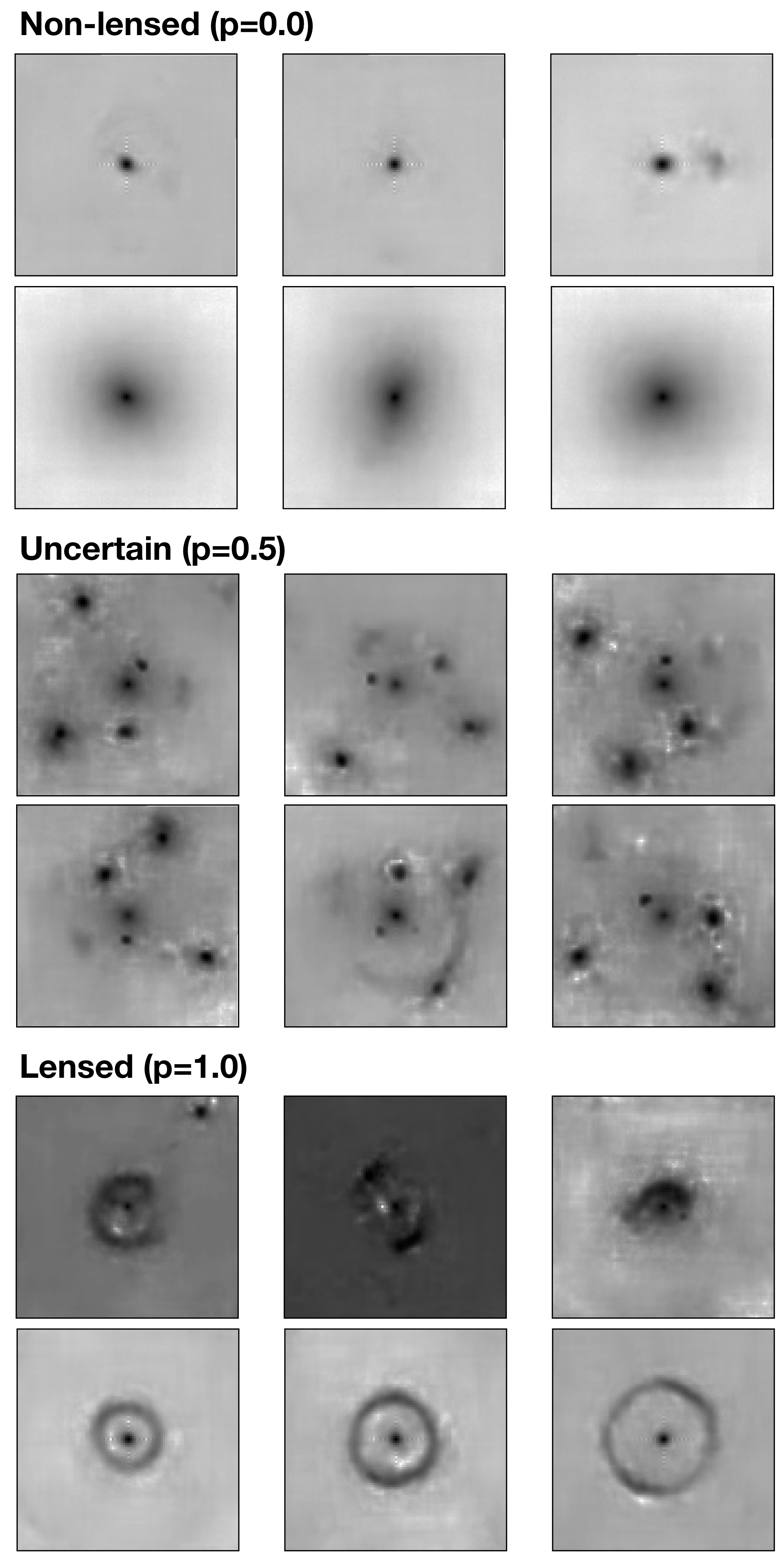}
   	\caption{Examples of the denoised images from which we assume the lensing probability for clusters. The `p' value represents the assumed lensing probability for clusters. {\it Top:} the examples of visually non-lensing images (p=0.0). {\it Middle:} the uncertain case (p=0.5). {\it Bottom:} the visually lensing images are presented (p=1.0).}
    \label{fig:example_assume_prob}
\end{center}
\end{figure}

Supervised machine learning methods applied to strong lens finding typically require tens of thousands of labelled images for training. This is of course too large for viable human classification and negates the whole purpose of using machine learning in the first place. Therefore, we propose a vastly streamlined way to calculate the predicted lensing and non-lensing probability for the $n$-th image by assuming the probability of each type for the $k$-th cluster through looking at the representative features of each cluster. We assume the lensing probability for the $k$-th cluster is 1.0, i.e. ${ P }_{ len }^{ k }=1.0$, if the representative features of this cluster have significant lensing features (e.g. Einstein rings, distorted arc, etc) (see the bottom of Fig.~\ref{fig:example_assume_prob}). If the features of this cluster are convincingly non-lensing features (e.g. singly isolated and oval object), the lensing probability of the $k$-th cluster is set to 0.0, i.e. ${ P }_{ len }^{ k }=0.0$ (see the top of Fig.~\ref{fig:example_assume_prob}). In the condition where it is difficult to classify such as those with multiple objects, the probability is assumed to be 0.5, i.e. ${ P }_{ len }^{ k }=0.5$ (see the middle of Fig.~\ref{fig:example_assume_prob}).

The summation of the lensing and non-lensing probabilities (equation~\ref{equ:probability}) may not be 1.0 when using assigned probabilities for clusters because the assigned probabilities cannot accurately represent the distribution of lensing and non-lensing images in each cluster. Therefore, we unify the predicted lensing and non-lensing probabilities as follows: ${ P }_{ len }^{ n' }={ { { P }_{ len }^{ n } } }/{ \left( { P }_{ len }^{ n }+{ P }_{ non }^{ n } \right)  }$ and ${ P }_{ non }^{ n' }={ { { P }_{ non }^{ n } } }/{ \left( { P }_{ len }^{ n }+{ P }_{ non }^{ n } \right)  }$.

The combination of assigned probabilities within our unsupervised technique promisingly reduces the quantitative effort of human judgement on data labelling whereby experts classify a few images that are grouped based on features rather than derived by a machine using over 10,000 images. The comparison of the results using true fractions and assumed probabilities are discussed in section~\ref{sec_comp_prob}. 
\subsection{Examinations}
\label{sec_exam}
\begin{figure}
\begin{center}
\graphicspath{}
	\includegraphics[width=0.6\columnwidth]{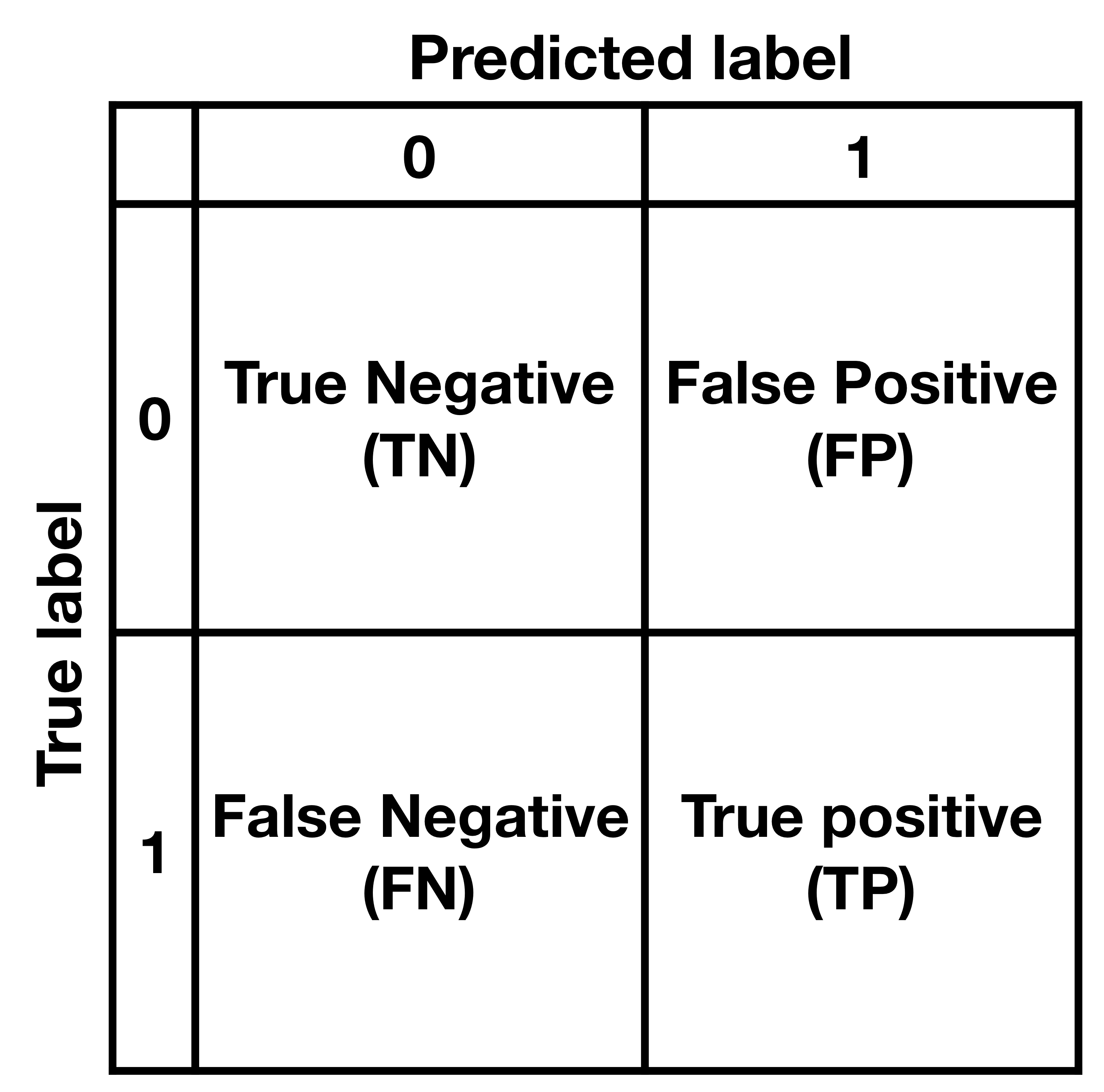}
   	\caption{The confusion matrix. The \textit{x}-axis label is the predicted label and the \textit{y}-axis label is the true label. The `0' means negative as well as non-lensing type while `1' represents positive signal and lensing type in this study.}
    	\label{fig:illusion_TP_FN}
\end{center}
\end{figure}

With the information on the lensing and non-lensing probability in each  cluster, we can compare the performance of our technique with other supervised machine learning techniques using the Receiver Operating Characteristic curve (ROC curve) \citep{Fawcett2006, Powers2011}. On a ROC curve the \textit{y}-axis is the true positive rate and the \textit{x}-axis is the false positive rate; therefore, the closer the ROC curve gets to the corner (0,1), the better the performance is. The definition of the true positive and the false positive are shown in Fig.~\ref{fig:illusion_TP_FN} in terms of the confusion matrix. Therefore, the true positive rate ($TPR$) and false positive rate ($FPR$) are defined as,
\begin{equation}
    	TPR=\frac { TP }{ TP+FN } ;\quad FPR=\frac { FP }{ FP+TN }.
\end{equation}
With the ROC curve, an evaluation factor called `area under the Receiver Operating Characteristic curve (AUC)' \citep{Bradley1997, Fawcett2006} is measured to evaluate the performance of machine learning algorithms. The AUC can be interpreted as the probability that a classifier ranks a randomly chosen positive example greater than a randomly chosen negative example. This factor also indicates the separability - how well the classifications can be correctly separated from each other.

\begin{figure}
\begin{center}
\graphicspath{}
	\includegraphics[width=\columnwidth]{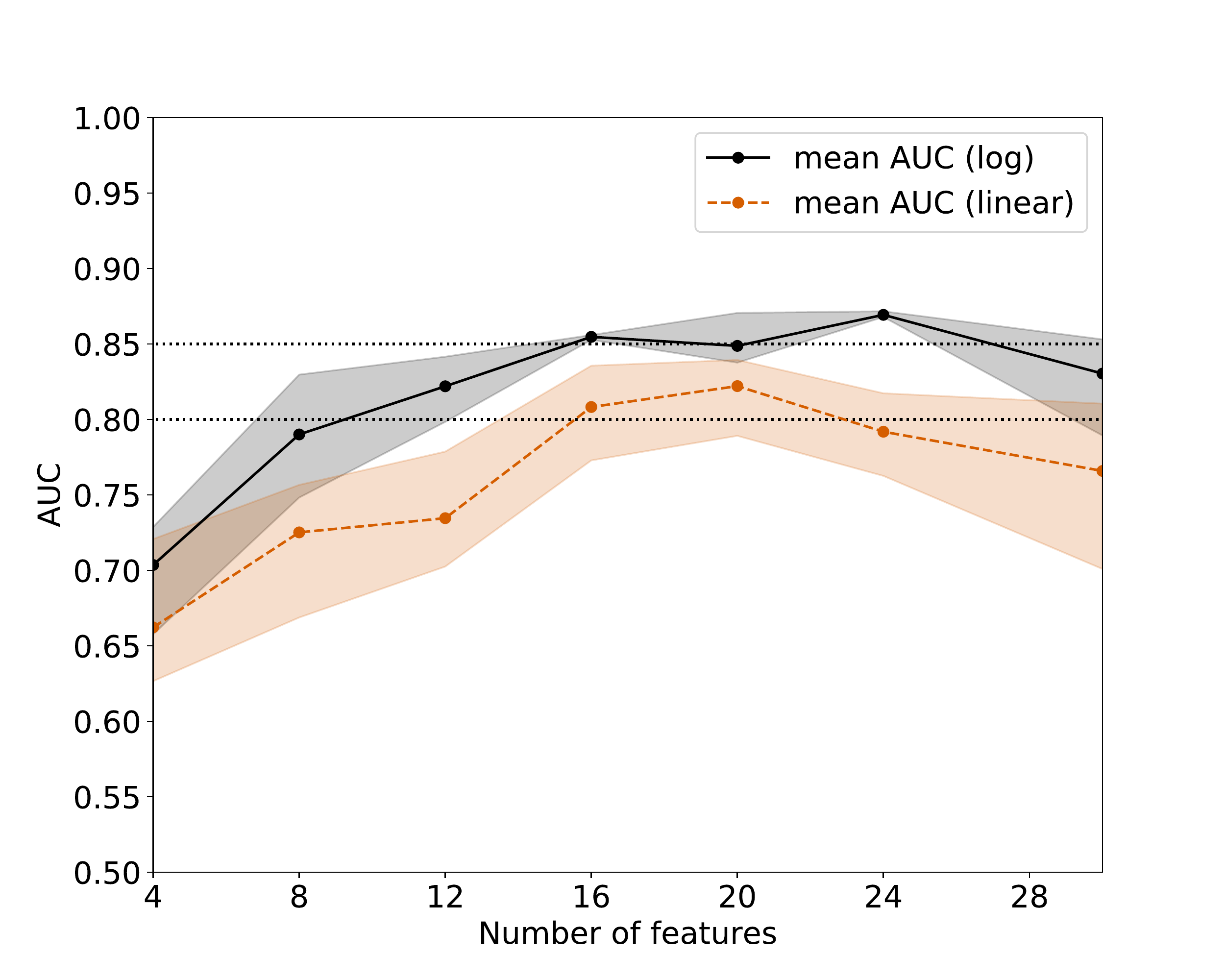}
   	\caption{The graph of AUC versus the number of extracted features in the CAE (Section~\ref{sec_cae}). The black solid line represents the mean value of the AUC trained by images with a logarithmic scale, and the orange dashed line is trained by images with a linear scale. The lighter shadings show the variation defined by the maximum and minimum of three reruns. The two dotted lines are locations of AUC = 0.80 and 0.85.}
    \label{fig:auc_features}
\end{center}
\end{figure}

In this study, we apply AUC to find the most optimal number of extracted features within the EL in the CAE. In Fig.~\ref{fig:auc_features}, the black solid line shows the results trained by the images in a logarithmic scale, and the lighter orange dashed line presents the one trained by the images within a linear scale. The lighter shadings show the variation in training defined by the maximum and minimum of three reruns.

Once the CAE model has been trained, the results of the clustering do not change as long as we use the same datasets. Therefore, the main uncertainty in the procedure is from the training process in the CAE. To determine the variation of results using different training we rerun our CAE three times for different numbers of features of the EL within the CAE, and use the maximal and minimal value of the AUC as the uncertainty for each number of features (Fig.~\ref{fig:auc_features}).

We discover that the CAE cannot reproduce the input images if we have an insufficient number of neurons in the EL. However, too many neurons cause overfitting such that the CAE captures noisy features. We find that the highest value of the AUC is carried out from the training by using logarithmically scaled images and the optimal number of neurons in the EL is 24 according to Fig.~\ref{fig:auc_features}. As such, we adopt this set up for all results presented in this work.

Apart from the ROC curve and the AUC value mentioned in section~\ref{sec_training}, we also use some other evaluation factors such as recall, precision, f1{\_}score, and accuracy, which are measured based on a probability threshold $p=0.5$. The definition of `recall' is identical to the $TPR$ in statistics which represents the completeness that shows the fraction of true types correctly identified, while `precision' indicates the contamination which means the fraction of true types in the list of candidates predicted. The `f1{\_}score' is a weighted average of the precision and recall which can be interpreted as the overall performance considering the contributions from both completeness and contamination. This is calculated by the formula \citep{Powers2011}:
\begin{equation}
    	{\rm f1}\quad =\quad 2\times \frac { \left({\rm precision}\times {\rm recall} \right)  }{ \left( {\rm precision}+{\rm recall} \right)}.
\end{equation}
The accuracy is defined by the formula:
\begin{equation}
\label{equ:accuracy}
    	{\rm accuracy}=\frac { TP+TN }{ TP+FP+TN+FN }, 
\end{equation}
such that the meaning of this is defined as how many successfully classified samples there are out of all the samples. 
\section{Results}
\label{sec_results}
In this section, we first compare the results using two different calculations of the lensing and non-lensing probabilities for each image (section~\ref{sec_clf}) in Section~\ref{sec_comp_prob}. The capability of our unsupervised technique to distinguish different types of lenses, and the performance of classification are presented in Section~\ref{sec_initial}. We also analyse our technique on the testing datasets with different fractions of lensing images; the result of this is shown in section~\ref{sec_frac}. Finally, we revisit the Strong Gravitational Lens Finding Challenge; we present our comparison with other supervised machine learning methods and human inspection in Section~\ref{sec_comp}.
\subsection{Comparison of Known and Assumed Probabilities}
\label{sec_comp_prob}
\begin{figure*}
\begin{center}
\graphicspath{}
	\includegraphics[width=2.1\columnwidth]{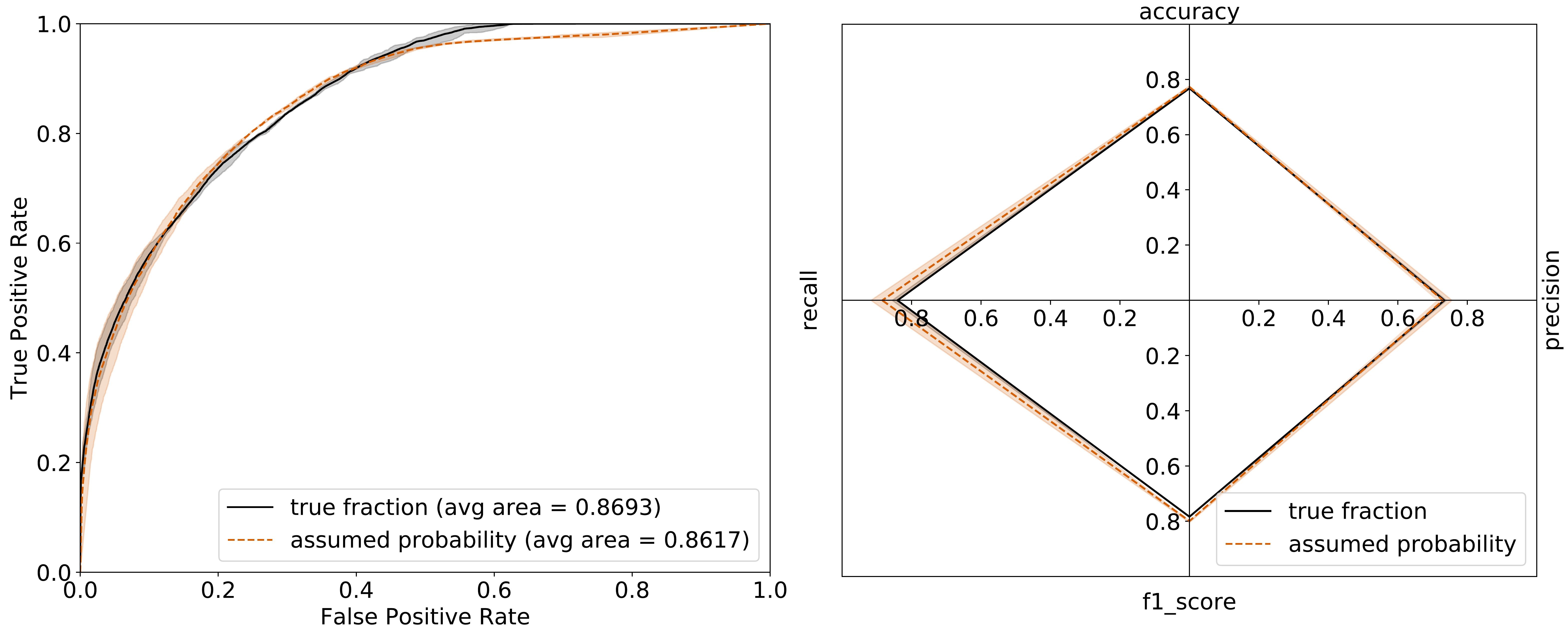}
   	\caption{The comparison of two methods to obtain the predicted probability of each class for each image using a known fraction and an assumed probability (section~\ref{sec_clf}). The black solid line represents the mean value using a known fraction, and the orange dashed line shows the mean value using an assumed probability of each class. The colour shadings are the variation defined by the maximum and minimum within three reruns. {\it Left}: the ROC curve. {\it Right}: the comparison of different statistic factors, e.g. recall, precision, f1{\_}score, accuracy.}
    \label{fig:comp_labelornot}
\end{center}
\end{figure*}

The comparisons of results with a known fraction of lensing and non-lensing images and an assumed probability of lensing (${ P }_{ len }^{ k }$) and non-lensing (${ P }_{ non }^{ k }$) in the $k$-th classification cluster (Section~\ref{sec_clf}) are shown in Fig.~\ref{fig:comp_labelornot} using images with logarithmic scale and 24 units in the embedded layer (EL) of the convolutional autoencoder (CAE). 

The left panel in Fig.~\ref{fig:comp_labelornot} presents the Receiver Operating Characteristic curve (ROC curve); the right panel is a comparison of different factors between these two methods such as recall, precision, f1{\_}score, and accuracy. In Fig.~\ref{fig:comp_labelornot}, the black solid line shows the mean value of the ROC curve using a known fraction of lensing images, and the orange dashed line represents the mean value of the results using an assumed probability. The colour shadings represent the variation defined by the maximum and minimum within three reruns.

Although the results of the `assumed probability' show larger scatter and slightly worse performance than the results of the `known fraction', the scatter of the `assumed probability' method is consistent with the results of the `known fraction' method. Additionally, the mean values of both methods are close to each other. Overall, these two methods show consistent results in their general performance, which is shown through the ROC curve, recall, precision, f1{\_}score, and accuracy (calculated based on a probability threshold of $p=0.5$).

This comparison confirms that the alternative calculation assigning an assumed probability to the classification clusters can be used to obtain promising lensing and non-lensing probabilities for each image. Furthermore, this indicates that the classification clusters obtained by our technique captures representative features from images and reflects the real lensing fractions in the clusters. Additionally, this result also shows an advantage of our technique for saving effort on data labelling by clustering the data before classifying it so that we can classify the feature of the small number of classification clusters instead of each image itself. This can be used as a preliminary selection method for future surveys when using a large amount of data.

\subsection{Identifying Lenses}
\label{sec_identify_lenses}
\subsubsection{Initial Results}
\label{sec_initial}
\begin{figure}
\begin{center}
\graphicspath{}
	\includegraphics[width=\columnwidth]{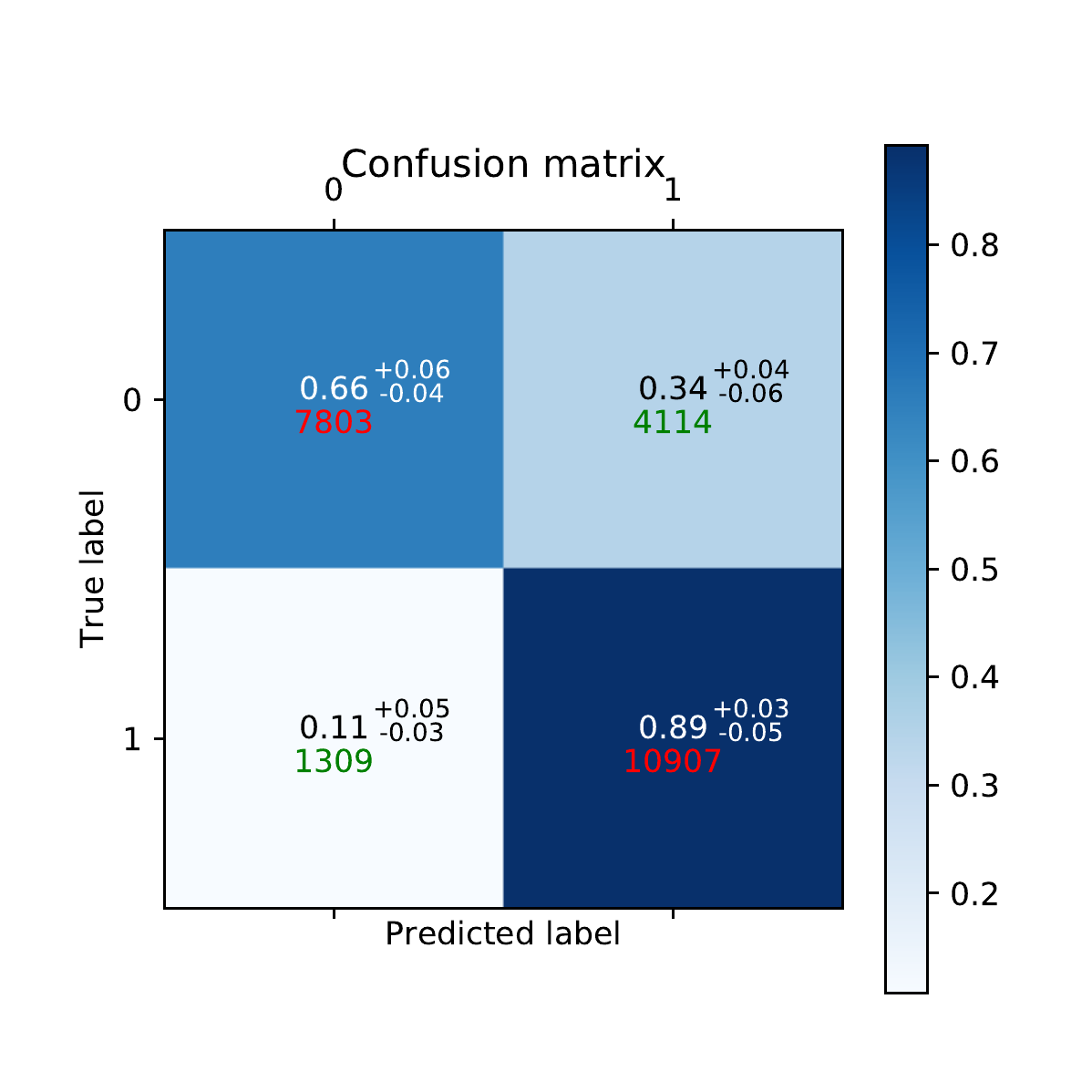}
   	\caption{The confusion matrix of the training set trained with 24 features in the embedded layer (EL) of the convolutional autoencoder (CAE). The floating values show the mean of the three reruns and the deviation from the maximum and minimum.} The red and green texts shown below the fraction are the actual number in the quadrant.
    	\label{fig:ex_training_r1}
\end{center}
\end{figure}

We begin with the results of binary classification using the predicted lensing probability obtained using the `assumed probability' method in Section~\ref{sec_clf}. In Fig.~\ref{fig:ex_training_r1}, we present the confusion matrix of the training set. The accuracy of our technique reaches $0.7725\pm 0.0048$ and the AUC reaches $0.8617\pm0.0063$ using a probability threshold of $p=0.5$. The error estimation of the accuracy on the AUC is based on the standard deviation of 3 reruns.

\begin{figure*}
\begin{center}
\graphicspath{}
	\includegraphics[width=1.5\columnwidth]{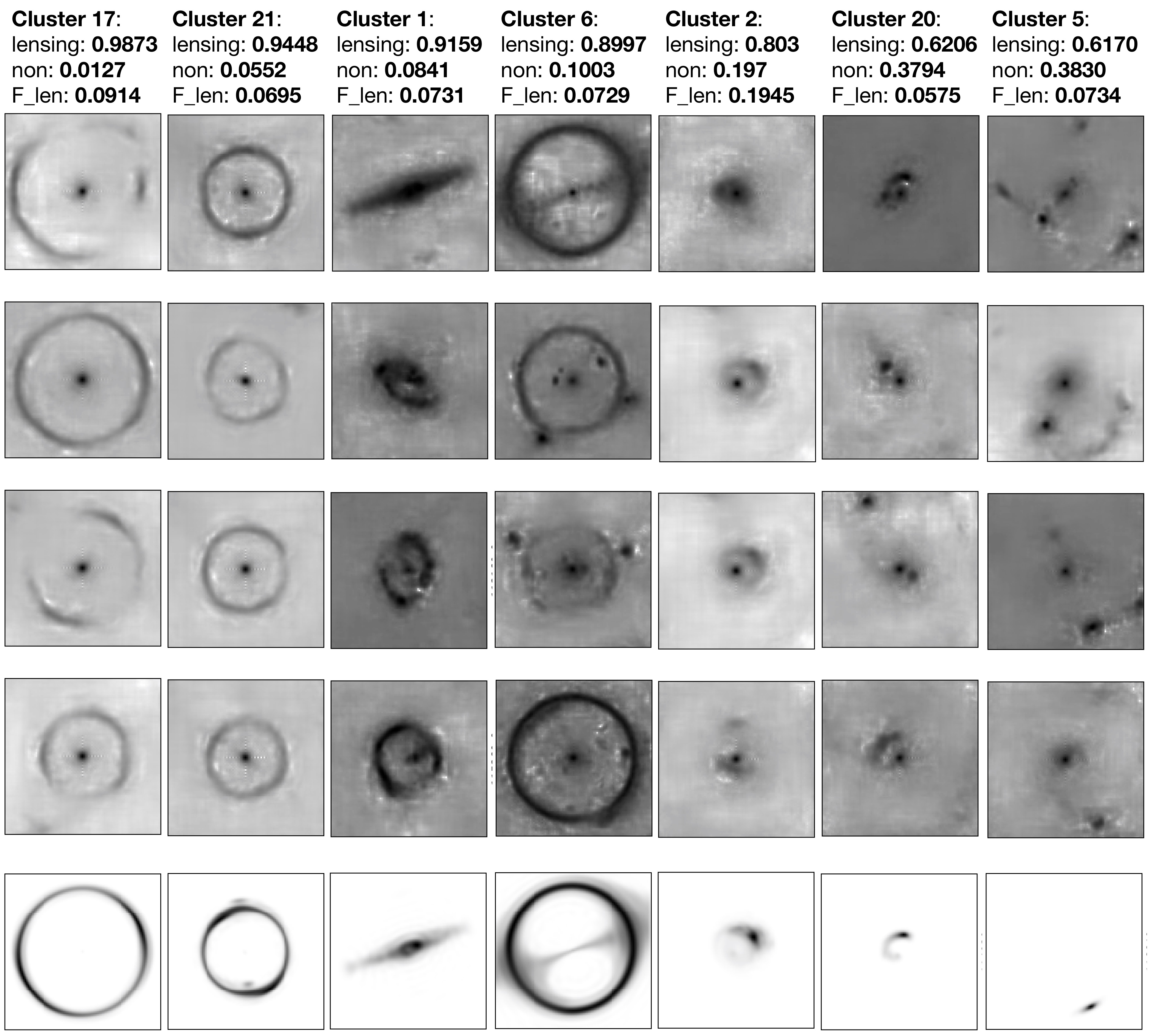}
   	\caption{Examples of the classification clusters having a high fraction of lensing types in individual clusters (denosied images). The top of each column shows the classification cluster index, the fraction of lensing (lensing) and non-lensing (non) in the cluster, and the fraction of lensing in the cluster of all lensing images in the training set (F\_len). The last row shows the simulated data without lenses within each column.}
    	\label{fig:examples_lenses}
\end{center}
\end{figure*}

This method promisingly separates features in a way similar to how a human would. Fig.~\ref{fig:examples_lenses} shows examples of the classification clusters with a high fraction of lensing images ($\ge$0.6). Every classification cluster shown in Fig.~\ref{fig:examples_lenses} has its own characteristic features, which indicates that our technique is able to capture the visual difference and similarity between images. Additionally, these classification clusters with a fraction of $\ge$0.6 contain $\sim$63 percent of lensing objects in the training set. The last row in Fig.~\ref{fig:examples_lenses} shows an example of the simulated data without lenses for the classification cluster. It is clear that our technique captures features such as Einstein rings with different radii, different strength, and distorted arc structures, etc, and images without lenses. The classification clusters with significant lensing features such as Einstein rings and arc structures are easily distinguishable (the fraction of lensing images in these groups is $\ge$0.8) in our results.

\begin{figure*}
\begin{center}
\graphicspath{}
	\includegraphics[width=2.1\columnwidth]{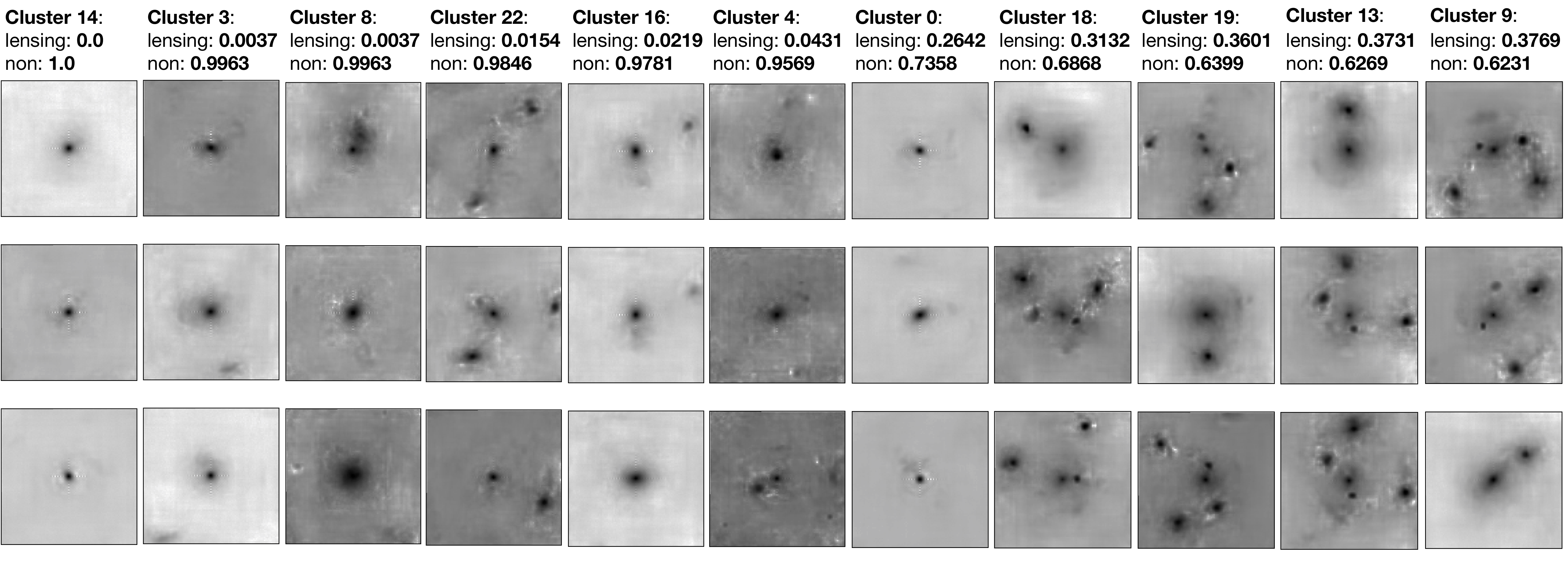}
   	\caption{Examples of the classification clusters having a high fraction of non-lensing images (denoised images). The top of each column shows the number of the cluster and the fraction of lensing (lensing) and non-lensing (non) in that cluster.}
    	\label{fig:examples_non-lensed}
\end{center}
\end{figure*}

In the same run, there are 7 classification clusters which have a high fraction of non-lensing images ($\ge$0.7); 6 out of 7 clusters include $\ge$0.9 fraction of non-lensing images. The features of these classification clusters are round or oval and isolated objects (Fig.~\ref{fig:examples_non-lensed}). The feature of cluster 0 looks oval and isolated, but has a relatively lower fraction of non-lensing images than others. It is produced by visually insignificant arc-like structures in the images that might also be created through the process of denoising. 

The last four columns in Fig.~\ref{fig:examples_non-lensed} which contain images with a fraction of non-lensing images between 0.6 and 0.7 are visually multiple objects. It is difficult to distinguish the classification of these types of images without colour information; however, our data is limited to a single visual band (section~\ref{sec_dataset}) so the decrease of performance is unavoidable. Additionally, these four classification clusters are similar to each other, but they are in a different orientation which shows that our technique cannot take care of rotation invariance at the current stage (also see Appendix~\ref{sec_nolenses} and the discussion in section~\ref{sec_discussion}). 

\begin{figure}
\begin{center}
\graphicspath{}
	\includegraphics[width=\columnwidth]{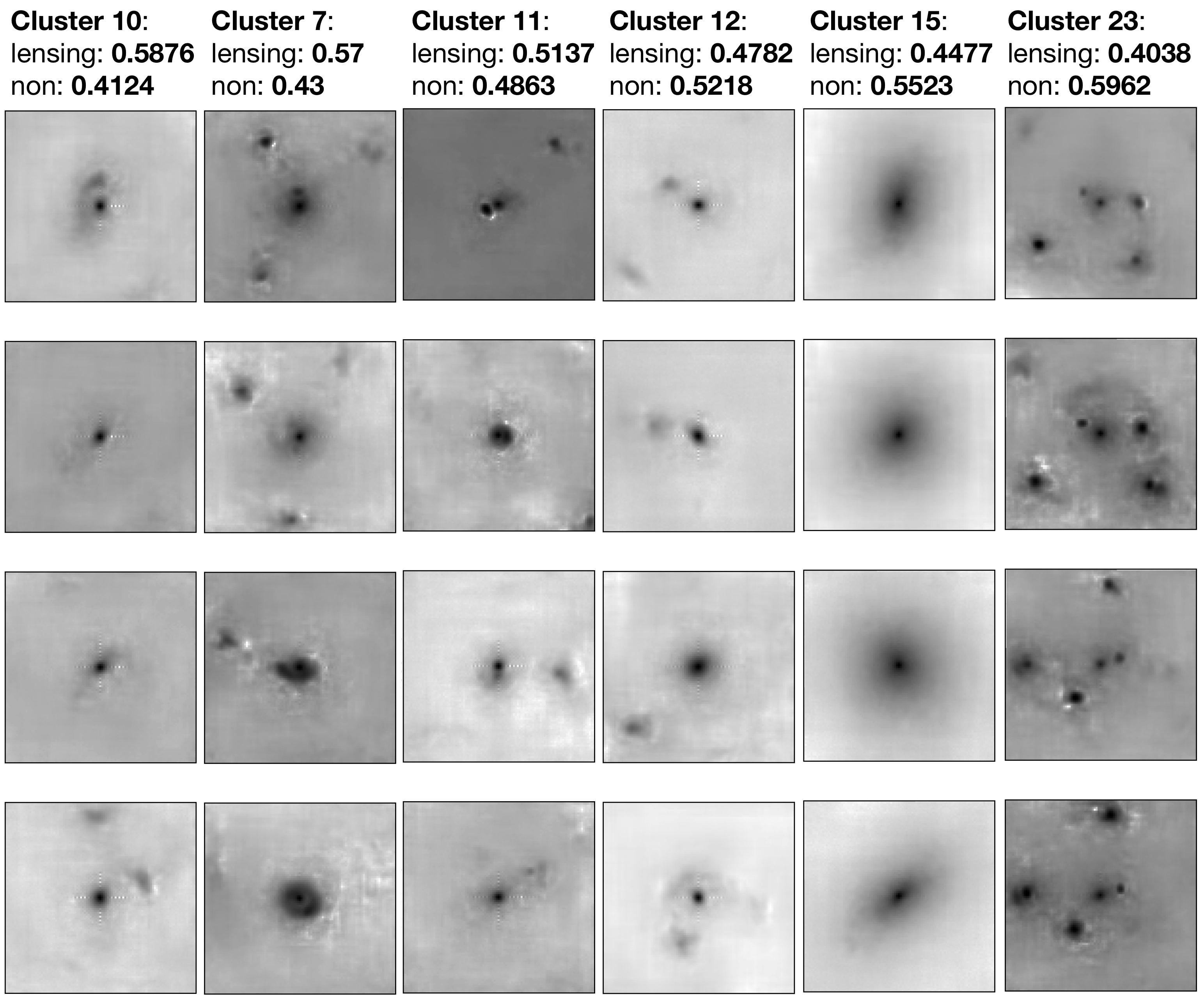}
   	\caption{Examples of the classification clusters with uncertain classification (denoised images). The top of each column shows the number of the classification cluster and the fraction of lensing (lensing) and non-lensing (non) in the cluster.}
    	\label{fig:examples_uncertain}
\end{center}
\end{figure}

The remaining 6 classification clusters are regarded as uncertain types because the fractions of lensing images in these groups are within the range from 0.4 to 0.6 (Fig.~\ref{fig:examples_uncertain}). Apart from clusters 15 and 23, the features of other classification clusters are single or double objects with filament or arc-like structures which might also be generated by the denoising process. The main features of cluster 15 is a round and single object with lenses surrounded by a halo-like structure, which can occur when the Einstein radius of lensing is equal to or smaller than the size of lenses. On the other hand, cluster 23 has similar features to clusters 9, 13, 18, and 19 which all show multiple object types in the images. As mentioned in the previous paragraph, the images shown in the clusters 15 and 23 cannot be easily distinguished without colour information; therefore their categories are ambiguous.

Overall, it is more challenging to correctly classify images of lensing and non-lensing types without significant lensing features, such as Einstein rings, and highly distorted arc structures seen using our technique with a single band. Our method obtains classification clusters with lensing features containing $\sim$63 percent lensed images from all lensed images in the training set (Fig.~\ref{fig:examples_lenses}). The remaining lensed images are distributed in the classification clusters with difficult features (e.g. the last four columns in Fig.~\ref{fig:examples_non-lensed} and Fig.~\ref{fig:examples_uncertain}).

We anticipate that the inclusion of colour will enhance the performance of this method on the basis that additional diagnostic information would be provided from other surveys with multiple broad-band filters rather than the single Euclid Space Telescope with VIS band.

As part of our investigation, we applied our pre-trained CAE on the simulated data without lenses (central galaxies) (Appendix~\ref{sec_nolenses}). Examples are shown in Fig.~\ref{fig:examples_nolenses} which confirms that the CAE promisingly captures the structure of different lensing types: Einstein rings with different radii, incomplete Einstein rings, arc structures with different lengths and positions, extended objects, etc, from these simulated images. 
\subsubsection{Test on datasets with different fractions of lenses}
\label{sec_frac}
A detectable galaxy-galaxy strong lensing event is an extremely rare event in the universe, e.g. 0.05 percent of 640,000 early type galaxies in the Canada France Hawaii Telescope Legacy Survey are strong galaxy-galaxy lenses \citep{Gavazzi2014}. To be capable of a more realistic case, we test our CAE and pre-trained Bayesian Gaussian mixture model (BGM) on datasets using logarithmic images with different fractions of lensing images from 50 percent to only 0.01 percent of lensing images \citep{Collett2015} (Table~\ref{tab:testing}).

\begin{figure}
\begin{center}
\graphicspath{}
	\includegraphics[width=\columnwidth]{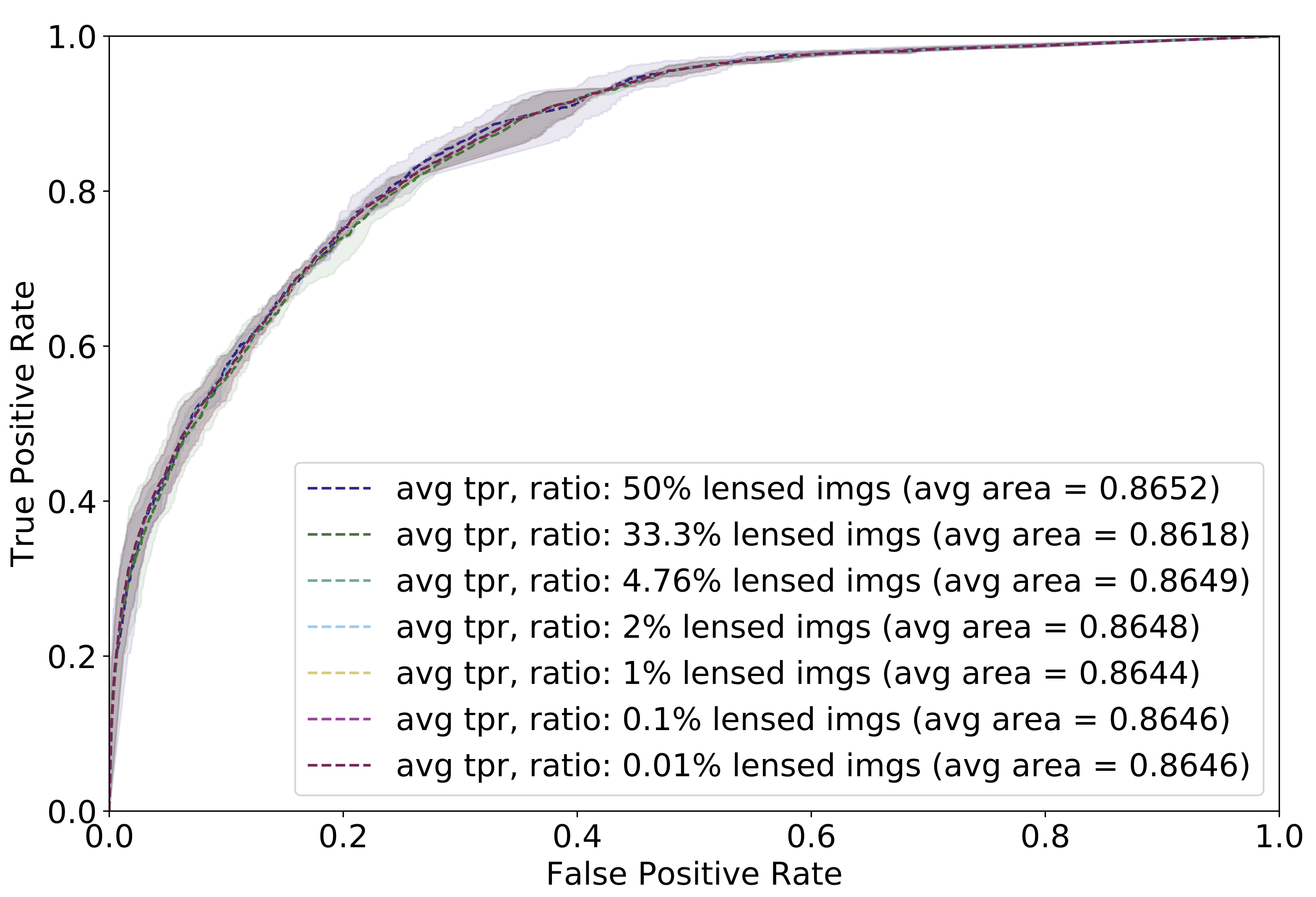}
   	\caption{The ROC curve of the testing sets using different fractions of lensing images. Different colours represent different fractions (Table~\ref{tab:testing}). The dashed lines show the average of the ROC curves within three reruns and the shading areas show the variation.}
    \label{fig:roc_ratios}
\end{center}
\end{figure}

The results are shown in Fig.~\ref{fig:roc_ratios}. Here we always use the `assumed probability' to calculate the predicted probability of each type for each image (section~\ref{sec_clf}). Different colours represent testing sets with different fractions of lensing and non-lensing images. The dashed lines are the average of the ROC curves and the shadings are the variation within three reruns. 

\begin{figure}
\begin{center}
\graphicspath{}
	\includegraphics[width=\columnwidth]{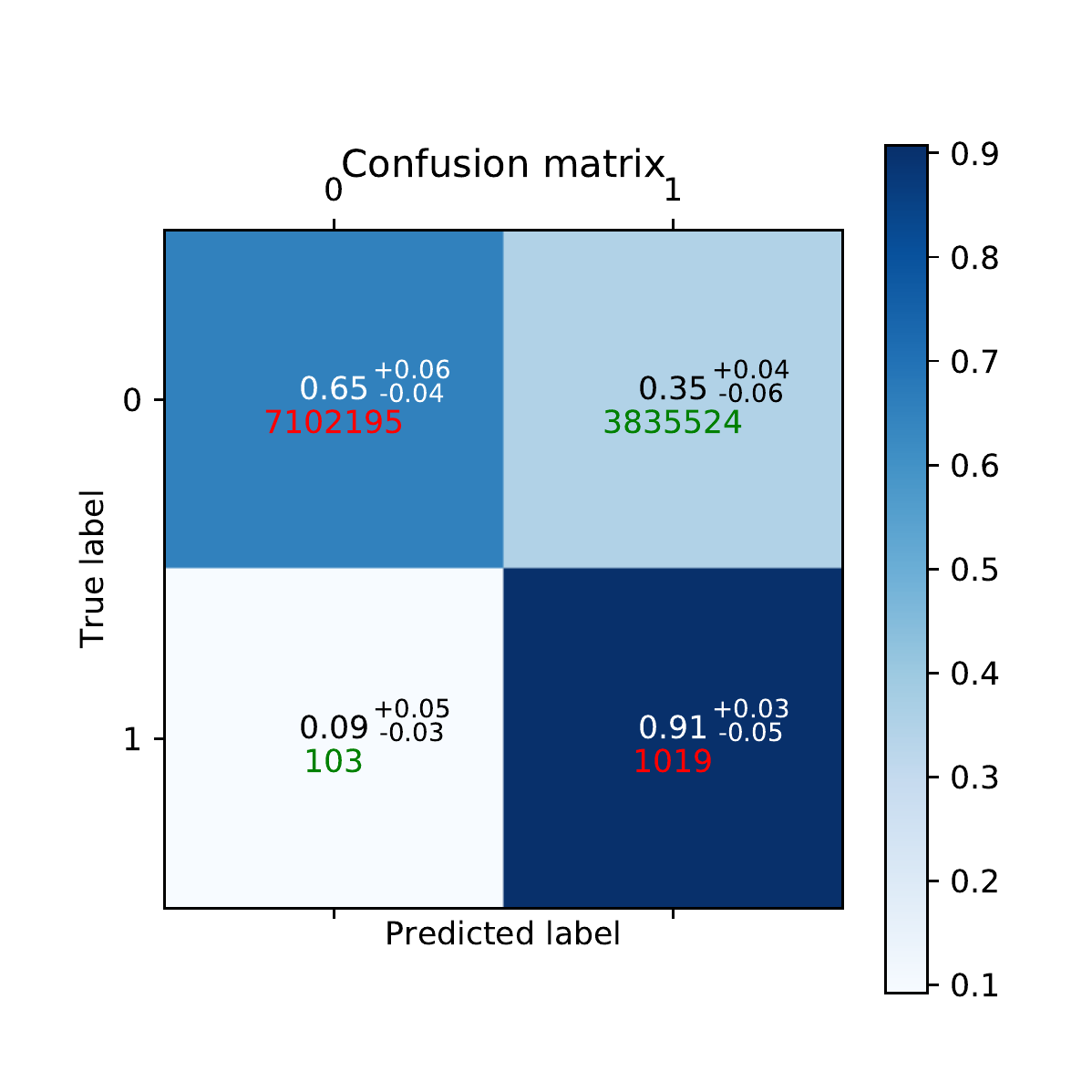}
   	\caption{The confusion matrix of the testing set containing 0.01 percent lensing images using the pre-trained model with 24 neurons in the embedded layer (EL) of the convolutional autoencoder (CAE). The floating values show the mean of the three reruns and the deviation from the maximum and minimum. The red and green texts shown below the fraction are the actual number in the quadrant.}
    \label{fig:example_r10000}
\end{center}
\end{figure}

Fig.~\ref{fig:roc_ratios} clearly shows that there is not a significant difference between the performance of the testing sets with different fractions of lensing images using our technique. Secondly, Fig.~\ref{fig:example_r10000} shows the accuracy of the classification in terms of a confusion matrix using the testing set with 0.01 percent of lensing images; this result is consistent with the results from training (Fig.~\ref{fig:ex_training_r1}). 

Both figures show that our unsupervised machine learning technique can maintain its performance even if the lensing events are rare in the data (to 0.01 percent of lensing images) when the model is well pre-trained.
\subsubsection{Comparison with Other Methods}
\label{sec_comp}
\begin{figure*}
\begin{center}
\graphicspath{}
	\includegraphics[width=2.1\columnwidth]{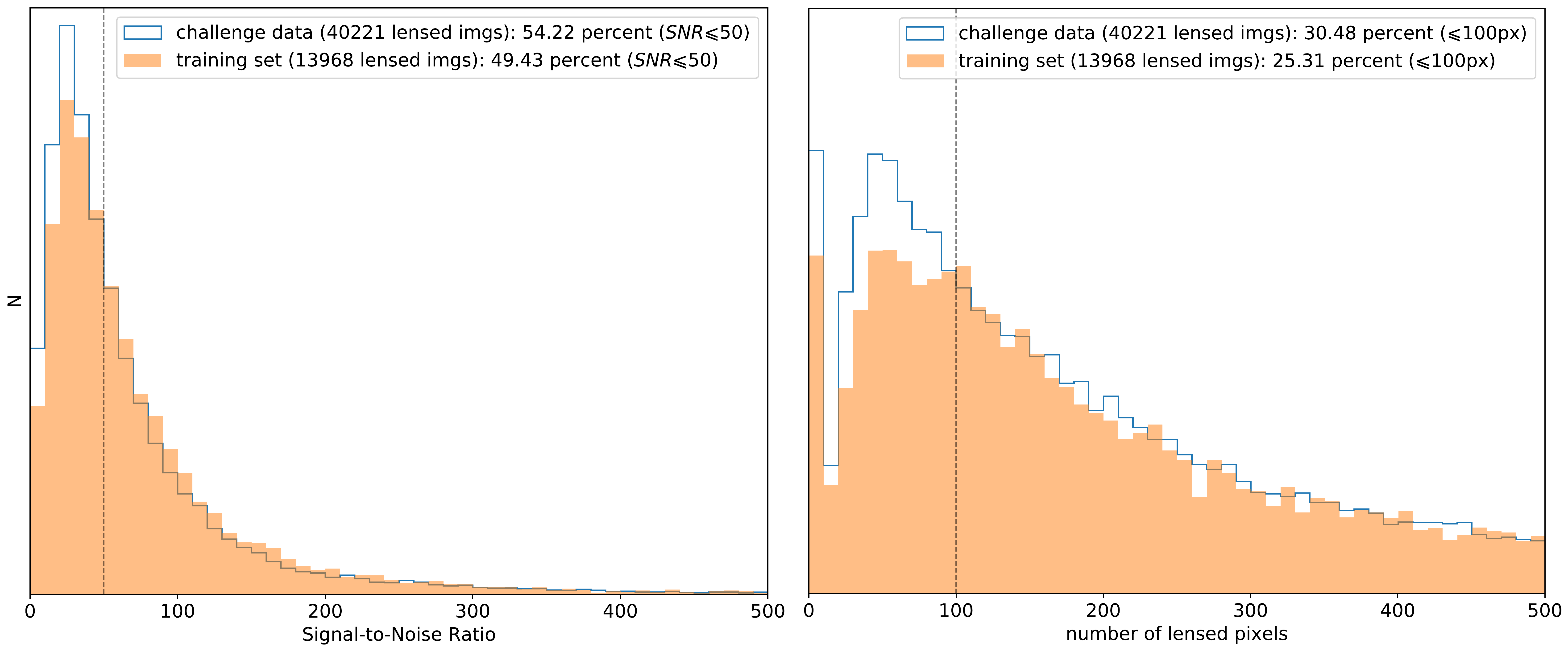}
   	\caption{The comparison of the Signal-to-Noise Ratios (SNR) and the number of lensed pixels above $1\sigma$ comparing the training set and the challenge testing data. {\it Left:} the comparison of SNR. {\it Right:} the comparison of the number of lensed pixels above $1\sigma$. The dashed lines represents the divide based on a visual assessment whereby the distribution on the left shows significant inconsistency between the training set and the challenge data set.}
    \label{fig:SNR_PIX}
\end{center}
\end{figure*}
To further compare the performance of our technique with other supervised machine learning methods and human inspection, we revisit the Strong Gravitational Lens Finding Challenge (Lens Finding Challenge) \citep{Metcalf2018}. The final challenge testing data in the Lens Finding Challenge includes 100,000 images, which are $\sim$60 percent of non-lensing images and $\sim$40 percent of lensing images. 

A visually detectable lensing feature generally has a high Signal-to-Noise Ratio (SNR) or has a low SNR but a larger number of correlated lensed pixels. Fig.~\ref{fig:SNR_PIX} shows the comparison of the SNR and the number of lensed pixels above $1\sigma$ between the training set and the challenge testing data. The value of the SNR in Fig.~\ref{fig:SNR_PIX} is calculated by $SNR=\frac { S }{ \sigma \sqrt { N }  } $, where $\frac { S }{ \sigma  } $ represents the intensity (flux) in a sigma contributed by the $N$ lensed pixels. This figure shows that the fraction of the images that are difficult to visually classify has increased from the training set to this challenge testing data.  

In addition to the value of AUC, \citet{Metcalf2018} apply two other factors: $TPR_0$ and $TPR_{10}$ to score the performance of their techniques. The $TPR_0$ is defined as the highest $TPR$ reached when the $FPR$=0 in the ROC curve. This quantity is used to recognise the classifiers whose highest classification levels are not conservative enough to eliminate all false positives; therefore, the $TPR_0$ of these classifiers are often equal to 0. The $TPR_{10}$ is defined when $TPR$ at the point where less than ten false positive are made. 

We apply the same architecture for the CAE as we do for the training set (Fig.~\ref{fig:cae}), followed by the training process shown in section~\ref{sec_training}, and the classifying process shown in section~\ref{sec_clf} whereby we are applying the `assumed probability' to this challenge testing data. The results are shown in Table~\ref{tab:comp}. 

Our unsupervised machine learning technique using a single band is more sensitive to significant lensing features. However, the challenge testing data contains the most visually difficult images with lower SNR and fewer lensed pixels resulting in poorer performance (`Unsupervised technique' in Table~\ref{tab:comp}) compared to the training set (labeled as * at the bottom row in Table~\ref{tab:comp}).

\begin{figure}
\begin{center}
\graphicspath{}
	\includegraphics[width=\columnwidth]{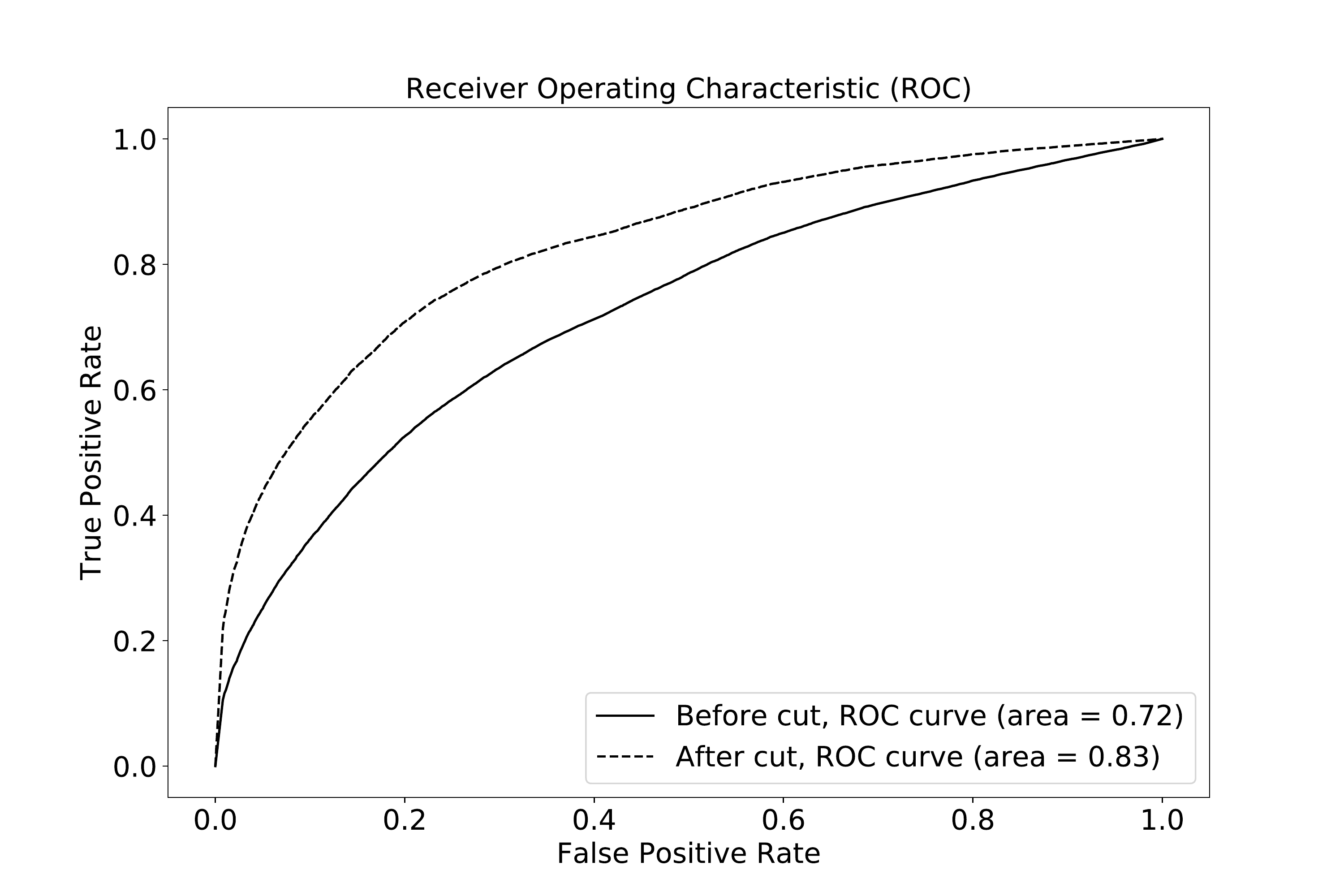}
   	\caption{The comparison of the ROC curve between before and after a cut at images with sizes greater than 100 lensed pixels and with a Signal-to-Noise Ratio larger than 50.}
    \label{fig:roc_between_cut}
\end{center}
\end{figure}

To fully test our method, we make a cut at 100 pixel and 50 SNR to exclude visually difficult images. This cut is determined by Fig.~\ref{fig:SNR_PIX} and a visual assessment to the images with these criteria. Applying this cut improves the performance of our technique from $AUC=0.72$ to $AUC=0.83$ that indicates that the difference in performance (i.e. AUC) between the two highlighted entries in Table~\ref{tab:comp} using our method is caused by the difference in the distribution of SNR and lensed pixels between the training and testing data. The comparison between applying the cut and not doing so is shown in Fig.~\ref{fig:roc_between_cut}. 

As in most methods, both $TPR_{0}$ and $TPR_{10}$ are equal to 0.00 using the challenge testing data in our results. However, in Fig.~\ref{fig:roc_between_cut}, both curves have a nearly vertical line at False Positive Rate $\sim$0 until True Positive Rate $\sim$0.1 (before) and $\sim$0.2 which means that although our technique is not able to eliminate all the misclassifications when the probability threshold is high (left), there are only a tiny number of images which were predicted incorrectly. 

This comparison gives an idea for the feasibility of this unsupervised machine learning technique compared with supervised methods. However, unsupervised machine learning is a qualitatively different method than supervised methods, such that unsupervised methods can explore data without label limitations and addresses questions that current supervised methods cannot. Therefore, the performance of unsupervised machine learning methods cannot simply be compared to supervised methods where the true label information is used.
\begin{table*}
\centering
\begin{tabular}{llrrrl}
  \hline
  Name & Author & AUC & $TPR_0$ & $TPR_{10}$ & short description \\ 
  \hline
  \hline
   LASTRO EPFL & Geiger, Sch\"{a}fer \& Kneib & 0.93 & 0.00 & 0.08 & CNN \\ 
   CMU-DeepLens-Resnet & Francois Lanusse, Ma, & 0.92 & 0.22 & 0.29 & CNN \\ 
   {} & C. Li \& Ravanbakhsh & {} & {} & {} & {} \\
   GAMOCLASS & Huertas-Company, Tuccillo, & 0.92 & 0.07 & 0.36 & CNN \\ 
   {} & Velasco-Forero \& Decenci\`{e}re & {} & {} & {} & {} \\ 
   CMU-DeepLens-Resnet-Voting & Ma, Lanusse \& C. Li & 0.91 & 0.00 & 0.01 & CNN \\ 
   AstrOmatic & Bertin & 0.91 & 0.00 & 0.01 & CNN \\ 
   CMU-DeepLens-Resnet-aug & Ma,  Lanusse, Ravanbakhsh & 0.91 & 0.00 & 0.00 & CNN \\ 
   {} & \& C. Li & {} & {} & {} & {} \\ 
   Kapteyn Resnet & Petrillo, Tortora, Kleijn, & 0.82 & 0.00 & 0.00 & CNN \\ 
   {} & Koopmans \& Vernardos & {} & {} & {} & {} \\ 
   CAST & Bom, Valent\'{\i}n \& Makler & 0.81 & 0.07 & 0.12 & CNN \\ 
   Manchester1 & Jackson \& Tagore & 0.81 & 0.01 & 0.17 & Human Inspection \\ 
   Manchester SVM & Hartley  \& Flamary & 0.81 & 0.03 & 0.08 & SVM / Gabor \\ 
   NeuralNet2 & Davies \& Serjeant & 0.76 & 0.00 & 0.00 & CNN / wavelets \\ 
   YattaLensLite & Sonnenfeld & 0.76 & 0.00 & 0.00 & Arcs / SExtractor \\ 
   All-now & Avestruz, N. Li \& Lightman & 0.73 & 0.05 & 0.07 & edges/gradiants and Logistic Reg. \\ 
   \bf Unsupervised technique & \bf This Work (Section~\ref{sec_comp}) & \bf 0.72 & \bf 0.00 & \bf 0.00 & \bf Deep Clustering \\
   GAHEC IRAP & Cabanac & 0.66 & 0.00 & 0.01 & arc finder \\ 
   \hline \hline
   \bf *Unsupervised technique & \bf This Work (Training, Fig.~\ref{fig:comp_labelornot}) & \bf 0.87 & \bf 0.08 & \bf 0.08 & \bf Deep Clustering \\   
   \hline
\end{tabular}
\caption{Edited based on the Table 3 in \citet{Metcalf2018}. The AUC, TPR$_0$ and TPR$_{10}$ for the entries in order of AUC. The highlighted entry without a * is the result of the challenge testing data (this Section). The bottom row with * shows the result obtained by using the training set (Fig.~\ref{fig:comp_labelornot}), which is used for comparing with the result of the testing data (the highlighted entry above without a *). The difference in AUC using our method between these two entries is due to the difference in the distribution of signal-to-noise ratio and lensed pixels between two datasets (Fig.~\ref{fig:SNR_PIX}).}
\label{tab:comp}
\end{table*}
\section{Future Work}
\label{sec_discussion}
In this paper, we describe an unsupervised machine learning technique for the detection of galaxy-galaxy strong gravitational lensing using simulated data based on the Euclid Space Telescope from the Strong Gravitational Lens Finding Challenge (Lens Finding Challenge) \citep{Metcalf2018}. This technique uses feature extraction provided by a convolutional autoencoder (CAE) and a Bayesian Gaussian mixture model (BGM) clustering algorithm.

This is an initial step in the use of convolutional autoencoders for astronomical unsupervised learning problems and as such there are many further explorations and improvements for this technique. For instance, there are other types of autoencoders e.g. variational autoencoder \citep{Kingma2013} for feature learning, and other kinds of clustering algorithms to explore the features and the properties of the obtained groups e.g. hierarchical clustering such as Agglomerative Hierarchical Clustering \citep{Bouguettaya2015} and density-based clustering such as DBSCAN \citep{Ester1996}, etc. 

In addition to other approaches that could be taken with different autoencoders and different clustering algorithms, some other future improvements are discussed here. First of all, we use the simulated data with a single VIS band in the optical region for the Euclid Space Telescope from Lens Finding Challenge. As shown in Section~\ref{sec_initial}, the lack of multiple bands causes difficulty in classifying certain types of images  (Fig.~\ref{fig:examples_uncertain}). In the future, we will apply our pipeline to surveys with multiple filters, which is expected to improve the performance further.

Secondly, the current state of this technique cannot preserve rotation invariance which means it categorises images differently when we rotate the images (see the last four columns in Fig.~\ref{fig:examples_non-lensed} {\&} Fig.~\ref{fig:examples_nolenses}). This condition does not affect the current results negatively in distinguishing lensing or non-lensing feature. However, considering the rotation invariance may help to reduce the number of classification clusters we obtain from this method when applying this technique on real data.

On the other hand, using an alternative autoencoder, the `variational autoencoder' \citep{Kingma2013} which applies Gaussian distributions to map the extracted features of each images is another potential approach to solve the issue of this rotation variance of clustering results. Preservation of rotation invariance in this way will be left for future work.

Thirdly, in our Appendix~\ref{sec_nolenses}, we show a perfect separation between lensing and non-lensing using the simulated data without lenses (i.e. central galaxies) within our technique. Although it is an unrealistic result considering we cannot perfectly deblend lenses and sources in real data, it is an indication of the improvement we might see without lenses through a pre-processing procedure of removing central galaxies.

One of the main issues of this technique is that we need a certain amount of data with strong features (e.g. lensed images, merger events, feature galaxies, etc) to let a CAE capture a variety of features from these objects. If the data with strong features is rare, the CAE would fail to capture the features and reproduce an inaccurate image.

The galaxy-galaxy strong lensing systems are relatively rare events in the universe. We have therefore had to use an amount of simulated data to train on. This situation could be potentially improved upon by further modification of the CAE architecture and possible data pre-processing. However, this technique is likely suitable for the astronomical objects with a relatively balanced distribution of features, such as the classification of galaxy morphology. However, few-shot learning \citep{Li2006} can be used when the labelled data is very limited. This could be one direction for improving the issue of having an extremely imbalanced data set within strong lensing detection scenarios.

On the other hand, the true power of an unsupervised machine learning technique is to find the hidden patterns or unrevealed characteristics in imaging data rather than just improving the efficiency or the performance for a known classification. To reveal the power of this unsupervised technique, we need to reconsider the selection method to determine the optimal number of the neurons in the embedded layer (EL) of the CAE to replace the value of AUC (Fig.~\ref{fig:auc_features}) in the future. Additionally, a forecast for the minimum number of features needed when using real observed data will be investigated in future work by improving the quality of the simulations and by adding more categories with realistic contamination. The ultimate determination for the optimal number of extracted features is also crucial for future usage when applying this unsupervised technique to observed data.


\section{Conclusion}
\label{sec_conclusion}
The purpose of this paper is to introduce an unsupervised machine learning technique that differs considerably from previous related works on the application to astronomical data. The unsupervised machine learning technique adopted in this paper is composed of the feature extraction by a convolutional autoencoder (CAE) and a clustering algorithm - a Bayesian Gaussian mixture model (BGM). We go beyond previous unsupervised work such as \citet{Hocking2018} and \citet{Martin2019} who applied Self-Organised Map (neural network) \citep{Kohonen1997} and hierarchical clustering to carry out feature extraction and clustering, respectively. 

We use the spaced-based simulated data from the Euclid Space telescope with a visual band (VIS) from the Strong Gravitational Lenses Finding Challenge (Lens Finding Challenge) \citep{Metcalf2018} and revisit this challenge. 

To compare our result with other lens-finding approaches, we propose a simple way to calculate the predicted probability of an image to be within each type - lensing and non-lensing by classifying the features of each cluster (Section~\ref{sec_clf}). This method, which promises to save an extensive effort need for data labelling in supervised machine learning, reaches an AUC value of $0.8617\pm0.0063$ and an accuracy of $0.7725\pm 0.0048$ on the classification of galaxy-galaxy strong lensing events using the training set of the space-based survey from the Lens Finding Challenge.

The main accomplishment of this study is that our technique captures meaningful features which follow human visual assessment from images without any initial label information. Additionally, this technique distinguishes a variety of lensing types (e.g. Einstein rings with different radii, different appearance of arcs) (Fig.~\ref{fig:examples_lenses} {\&} Fig.~\ref{fig:examples_nolenses}) and potentially can detect unusual lensing features. The discriminating ability is highlighted in Appendix~\ref{sec_nolenses} using a pre-trained CAE model on the simulated data without lenses.  

We then revisit the Lens Finding Challenge by applying our technique on their challenge testing data (section~\ref{sec_comp}). The results show a degradation in performance from the training set to the challenge testing data which is due to the difference in the distribution of the Signal-to-Noise Ratios (SNR) and the number of lensed pixels above $1\sigma$ in the lensed images in the challenge testing data. Therefore, we applied a cut at 100 pixels and 50 SNR to the challenge testing data, with the results shown in Fig.~\ref{fig:SNR_PIX}.  As can be seen, by removing these systems we improve the performance of our technique.

Another advantage of our technique is that it also retains its discriminating ability when the fraction of lensing images varies. As is shown in Section ~\ref{sec_frac}, the performance is consistent for the cases of the data holding $\sim$0.01 percent or $\sim$50 percent of lensing images, once the unsupervised model is well pre-trained.

The most promising advantage of this technique is the pre-selection in the process of searching for strong lenses in upcoming large scale imaging surveys. It reduces the sample size of the dataset needed for the classification by cleaning up apparent non-lensing systems. Also, our approach can identify rare lensing systems with unusual characteristics such as multiple Einstein Rings, which can be identified as non-lenses with a high probability by supervised finders if the training sets do not contain these features. 

In the future, as discussed in Section~\ref{sec_discussion}, we will try to improve the competitiveness of our approach by adopting different architectures of neural networks, alternative autoencoders or clustering algorithms. Combining unsupervised and supervised techniques is another direction we plan for increasing the performance of the identification of strong lenses. Finally, the development of a quantitative validation tool for unsupervised machine learning techniques such as the Receiver Operating Characteristic curve (ROC curve) \citep{Fawcett2006, Powers2011} for supervised machine learning techniques is of great importance for future work. Without such diagnostics, it is not possible to objectively compare unsupervised machine learning approaches.
\section*{Acknowledgements}
The authors acknowledges the support by the UK Science and Technology Facilities Council (STFC). Simon Dye is supported by a UK STFC Rutherford Fellowship. Ting-Yun Cheng gives a thank to the support of the Vice-Chancellor's Scholarship from the University of Nottingham, and discussions with Bobby Clement. 

\bibliographystyle{mnras}
\bibliography{ms} 

\appendix
\section{A Test on Simulated Data without Lenses}
\label{sec_nolenses}
\begin{figure*}
\begin{center}
\graphicspath{}
\includegraphics[width=2.1\columnwidth]{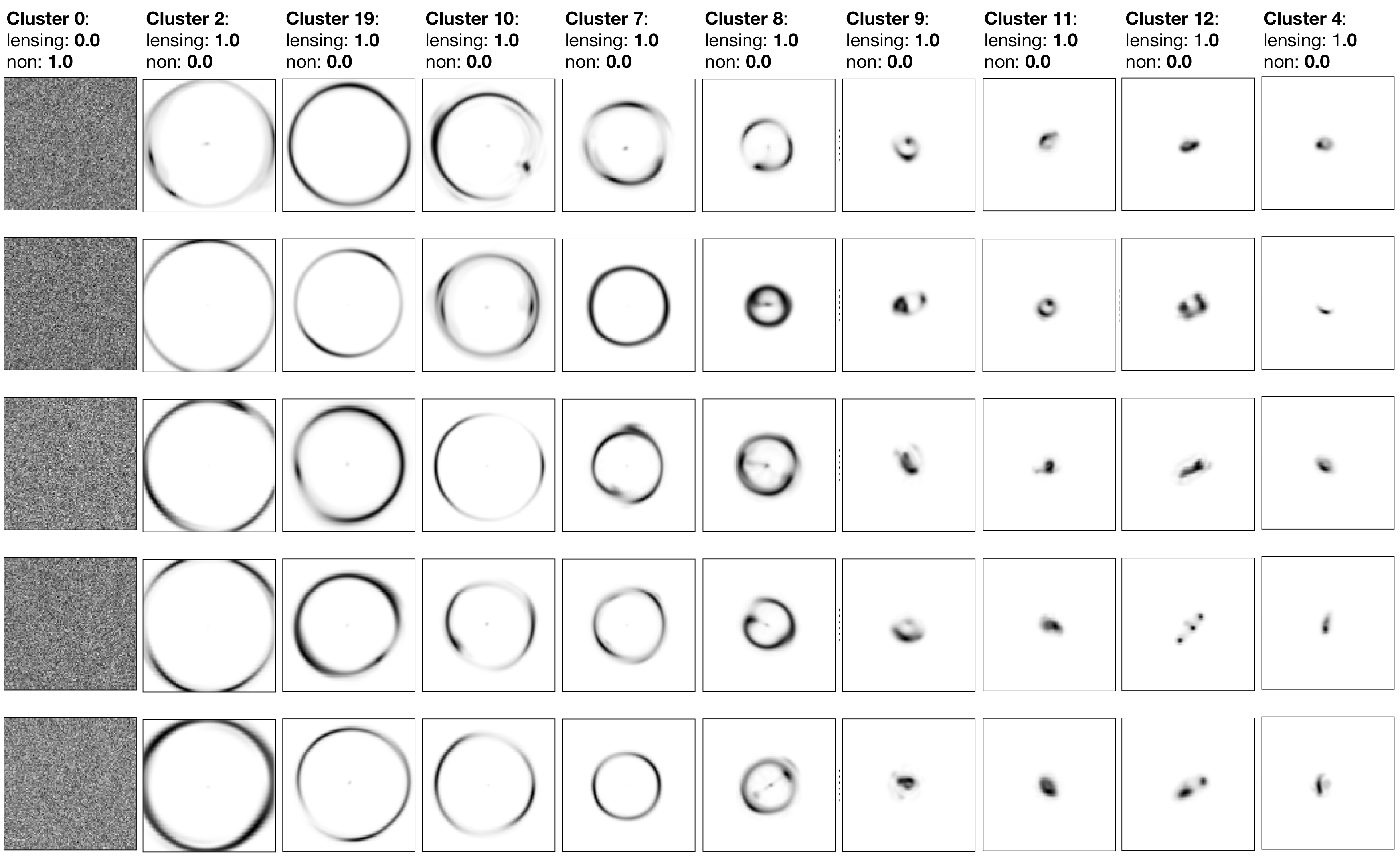}
   	\caption{Examples of classification clusters using the simulated data without lenses (central galaxies). The top of each column shows the number of the cluster and the fraction of lensing (lensing) and non-lensing (non) in the cluster. The figure is continued in Fig.~\ref{fig:examples_nolenses_2}.}
    \label{fig:examples_nolenses}
\end{center}
\end{figure*}
\renewcommand{\thefigure}{A1 (continued)}
\addtocounter{figure}{-1}
\begin{figure*}
\begin{center}
\graphicspath{}
	\includegraphics[width=2.1\columnwidth]{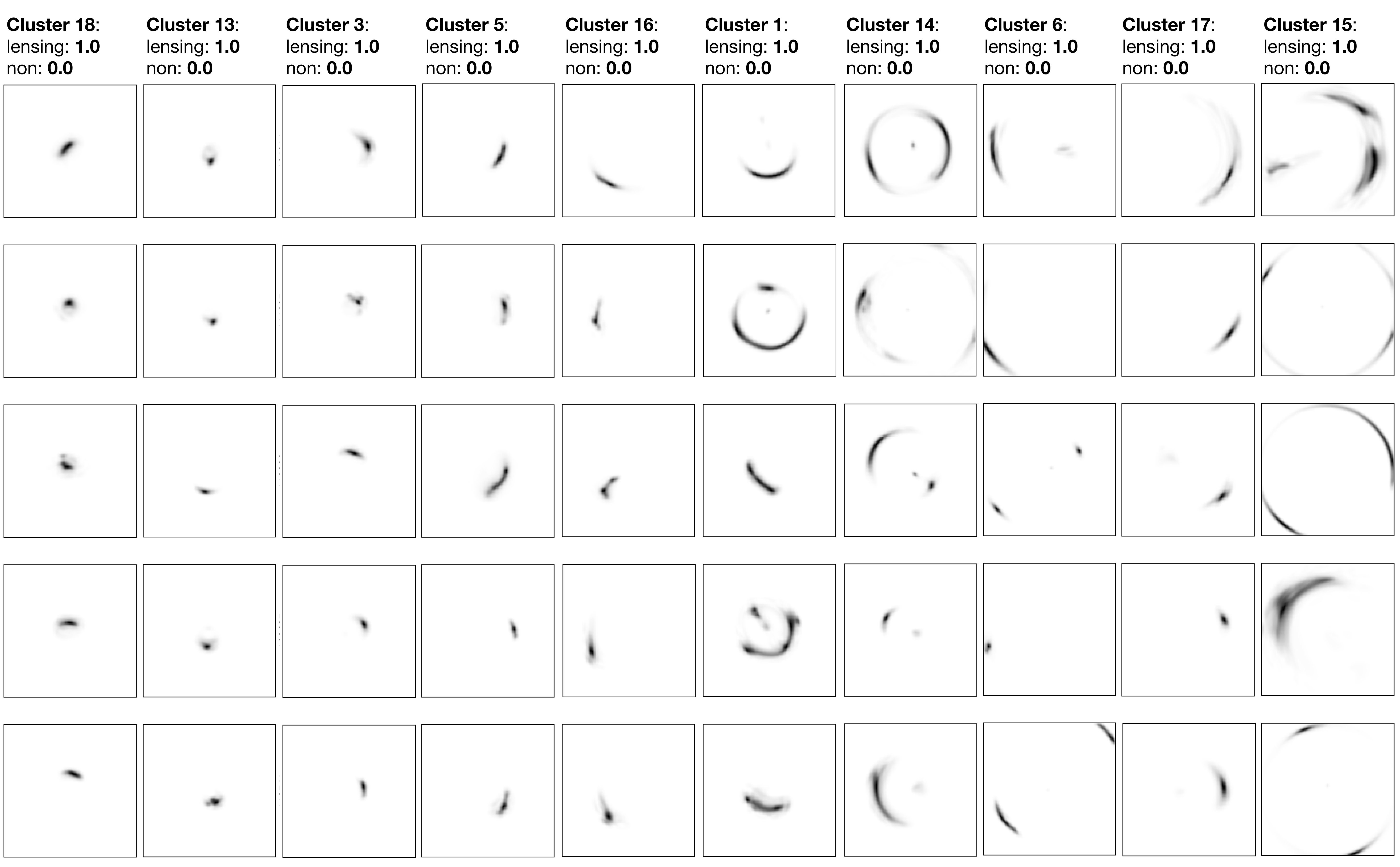}
   	\caption{The continued figure of Fig.~\ref{fig:examples_nolenses}.}
  	\label{fig:examples_nolenses_2}
\end{center}
\end{figure*}
\renewcommand{\thefigure}{\arabic{figure}}
As part of our investigation, we test our pre-trained convolutional autoencoder (CAE) (section~\ref{sec_training}) on our simulated data without lenses (i.e. central galaxies) in this study. The result is shown in Fig.~\ref{fig:examples_nolenses}. The purpose of this test is to explore the potential usefulness for this technique when deblending of the lenses from the sources is possible. 

The simulated data we used is the training set from the Strong Gravitational Lenses Finding Challenge (Lens Finding Challenge) \citep{Metcalf2018}. This challenge offered participants images with all possible image types (lenses, sources, and background noise), images with lenses only, and images with sources only. The simulated data without lenses (central galaxy, i.e. with source only) emphasizes the features of the images, thus, we use the pre-trained model trained by images with linear scale using 20 features (Fig.~\ref{fig:auc_features}) in the embedded layer (EL) of the CAE. 

The result reconfirms our results in section~\ref{sec_initial}. We ordered the clusters based on the appearance of the images in the cluster in Fig.~\ref{fig:examples_nolenses} such that it is easier to see the trend. Above the first row in Fig.~\ref{fig:examples_nolenses} shows the cluster ID and the fraction of both lensing (lensing) and non-lensing (non) in the cluster.

The first column (cluster) contains all the non-lensing images, which are shown as empty images when there are no lenses in the images. From the second to the eighth column in Fig.~\ref{fig:examples_nolenses} show the structure of Einstein rings with different radii and from the ninth column in Fig.~\ref{fig:examples_nolenses} to Fig.~\ref{fig:examples_nolenses_2} show the arcs structure with different features such as positions, lengths, or the radii of arcs.

We also reconfirm that the rotation invariance cannot be preserved using our current technique (the last four columns of Fig.~\ref{fig:examples_non-lensed} in section~\ref{sec_initial}). The characteristic of the CAE is to minimize the difference between input and output images; therefore, arcs with similar radii and lengths but located at different positions are identified as different clusters by our unsupervised technique at the current stage. Although this rotation variant has no significant effect on the final result, the improvement on considering rotation invariance might be helpful to reduce the complexity of extracted features when applying this technique to real data. 

Additionally, the lensing and non-lensing images are perfectly separated in this test. Although it is unrealistic, we might be able to significantly improve the performance and strengthen the usefulness of this technique by approaching the condition of the images in this test through a pre-processing procedure of removing central galaxies which is possible. 

\label{lastpage}
\end{document}